\documentclass[article]{revtex4-2}
\usepackage{amssymb}
\usepackage{epsfig}
\usepackage{units}
\usepackage{color}

\begin{document}

\large
\bigskip

\title{Measurement of the Generalized Polarizabilities of the Proton \\
in Virtual Compton Scattering \\
\qquad \smallskip \\
(VCS-II)
\qquad \smallskip \\
\qquad \smallskip \\
Proposal to Jefferson Lab PAC-51\\ }

\author{ \qquad  \\
\large H.~Atac~(spokesperson), R.~Li, N. Sayadat, S.~Shesthra, N.~Sparveris~(spokesperson / contact person\footnote{sparveri@temple.edu}), S.~Webster\\
\it{\small Temple University, Philadelphia, PA, USA}\\
\qquad \smallskip \\
\large A.~Camsonne~(spokesperson), J-P.~Chen, S.~Covrig-Dusa,  A.~Deur, \\ M.~K.~Jones~(spokesperson), M.~D.~McCaughan, A.~Tadepalli\\
\it{\small Thomas Jefferson National Accelerator Facility, Newport
News, VA, USA}\\
\qquad \smallskip \\
\large W.~Armstrong, S.~Joosten, Z.E.~Meziani, C.~Peng\\
\it{\small Argonne National Laboratory, Lemont, IL 60439, USA}\\
\qquad \smallskip \\
\large M.~Ali, M.~Paolone (spokesperson)\\
\it{\small New Mexico State University, NM, USA}\\
\qquad \smallskip \\
\large A.T. Katramatou, G.G. Petratos\\
\it{\small Kent State University, Kent, OH 44242}\\
\qquad \smallskip \\
\large E.~Brash\\
\it{\small Christopher Newport University, VA, USA}\\
\qquad \smallskip \\
\large J.~Bernauer, E.~Cline, W.~Li\\
\it{\small Stony Brook University, NY, USA}\\
\qquad \smallskip \\
\large G.~Huber\\
\it{\small University of Regina, Canada}\\
\qquad \smallskip \\
\large Z.~Zhao\\
\it{\small Duke University and Triangle Universities Nuclear Laboratory, NC 27708, USA}\\
\qquad \smallskip \\
\large N.~Liyanage, M.~Nycz\\
\it{\small University of Virginia, VA 22904, USA}\\
\qquad \smallskip \\
\large R.~Gilman\\
\it{\small Rutgers, The State University of New Jersey, Piscataway, New Jersey 08854, USA}\\
\qquad \smallskip \\
\newpage
\large C. Ayerbe Gayoso\\
\it{\small William and Mary, Williamsburg, VA 23185, USA}\\
\qquad \smallskip \\
\large M. Mihovilovi\v{c}, S. \v{S}irca\\
\it{\small Faculty of Mathematics and Physics, U. of Ljubljana \& 
Jo\v{z}ef Stefan Institute, Ljubljana, Slovenia}\\
\qquad \smallskip \\
\large M.~Demirci\\
\it{\small Karadeniz Technical University, Turkey}\\
\qquad \smallskip \\
\large D.~Androi\v{c}\\
\it{\small University of Zagreb, Faculty of Science, Croatia}\\
\qquad \smallskip \\
\large P. Markowitz\\
\it{\small Florida International University, Miami, FL 33199, USA}\\
\qquad \smallskip \\
\large A. Arora, D. Ruth, N. Santiesteban K. Slifer, A. Zec\\
\it{\small University of New Hampshire, NH, USA}\\
\large \color{white}{.}\\
\vspace{1.0in}
}

\date{\today}


\begin{abstract}

\end{abstract}

\maketitle


\newpage

\centerline {\Large \bf Executive Summary}
\vskip 0.4cm
\begin{description}
\item[Main Physics Goals:] The proposal focuses on the measurement of the electric and the magnetic 
Generalized Polarizabilities (GPs) of the proton~\cite{gpreview}.
\item[Proposed Measurement:] In Hall C, absolute cross section and cross section azimuthal asymmetry measurements for the $p(e,e'p)\gamma$ reaction will be conducted, spanning $Q^2=$ 0.05 to 0.50 (GeV/c)$^2$. The experiment will acquire production data for 59 days and 3 days for optics, normalization and dummy measurements for a total of 62 days of data taking. 
The electric and the magnetic GPs will be extracted from fits to the cross section and the asymmetry measurements. 
\item[Specific requirements on detectors, targets, and beam:] The HMS will detect protons using the standard detector package. The HMS will run at momentum between 494~MeV/c to 993~MeV/c and angles of 11.3$^{\circ}$ to 56.5$^{\circ}$. The SHMS will detect electrons using the standard detector package. The Noble gas cerenkov detector will be removed and replaced with the existing vacuum extension. The previous VCS experiment was run in this configuration.  The SHMS will run at momentum between 736.3~MeV/c to 1783~MeV/c and angles of 11.2$^{\circ}$ to 20.5$^{\circ}$.  A beam energy of 1.1 GeV ($\pm$~0.2 GeV) is needed for the lowest $Q^2$ kinematics and 2.2 GeV ($\pm$~0.2 GeV) for the rest of the kinematic settings, while the beam can be unpolarized.  The targets will be a 10-cm long liquid hydrogen, 10-cm aluminum dummy and optics foil targets. Elastic $ep$ coincidence is needed as measurement of HMS trigger efficiency and check on the HMS momentum optics. For these measurements, the HMS will be at angles of 50$^{\circ}$ to 63$^{\circ}$ and at momentum between 495 to 994~MeV/c while the SHMS will be at angles of  19$^{\circ}$ to 33$^{\circ}$ and at momentum between 924 to 1797~MeV/c.  
\item[Previous proposal:] This proposal aims to extend the work of experiment E12-15-001 (VCS-I). It follows the results and conclusions of the E12-15-001 that were recently published in~\cite{gp23}. It also follows the recommendation of the PAC44 report, namely to return to the PAC for additional beam time, once we have demonstrated that the systematics of these measurements are under control. We have submitted the VCS-II proposal for the continuation of the VCS program at JLab, aiming to pin down the dynamic signature of the proton GPs through high precision measurements combined with an extended and fine mapping as a function of $Q^2$.\end{description}.
\newpage

\quad \\
\quad \\
\quad \\
\quad \\

\qquad \qquad \qquad \qquad \qquad \qquad \qquad \qquad  {\bf Abstract} \\


We propose to conduct a measurement of the Virtual Compton
Scattering reaction in Hall C that will allow the precise extraction of the
two scalar Generalized Polarizabilities (GPs) of the proton in the region
of $Q^2=0.05~(GeV/c)^2$ to $Q^2=0.50~(GeV/c)^2$. The Generalized
Polarizabilities are fundamental properties of the proton, that characterize the system's response to an external electromagnetic (EM) field. They describe how easily the charge and magnetization distributions inside the system are distorted by the EM field, mapping out the resulting deformation of the densities in the proton. As such, they reveal unique information regarding the underlying system dynamics and provide a key for decoding the proton structure in terms of the theory of the strong interaction that binds its elementary quark and gluon constituents together. Recent measurements of the proton GPs have challenged the theoretical predictions, particularly in regard to the electric polarizability. The magnetic GP, on the other hand, can provide valuable insight to the competing paramagnetic and diamagnetic contributions in the proton, but it is poorly known within the region where the interplay of these processes is very dynamic and rapidly changing.   

The unique capabilities of Hall C, namely the high resolution of the
spectrometers combined with the ability to place the spectrometers
in small angles, will allow to pin down the dynamic signature of the GPs through high precision measurements combined with a fine mapping as a function of $Q^2$. The experimental setup
utilizes standard Hall C equipment, as was previously employed in the VCS-I (E12-15-001) experiment, namely the HMS and SHMS
spectrometers and a 10~cm liquid hydrogen target. A total of 59 days of unpolarized 75~$\mu A$ electron beam with
energy of 1100 MeV (6 days) and 2200 MeV (53 days) is requested for this experiment, along with three additional days for calibrations.
\\
\\

\newpage

\section{Introduction}

The proposed experiment offers to explore the scalar Generalized
Polarizabilities~\cite{gpreview} of the proton through measurements of the Virtual
Compton Scattering reaction in Hall C. The experiment will provide
high precision measurements of $\alpha_E$ and $\beta_M$ in the
region of $Q^2=0.05~(GeV/c)^2$ to $Q^2=0.50~(GeV/c)^2$. These
measurements will contribute in an instrumental way towards a deeper
understanding of the nucleon dynamics. 

The new proposal follows the steps of experiment E12-15-001 (VCS-I) that acquired data in year 2019. The results of this experiment were recently published in~\cite{gp23}. They have offered the most precise measurement of the proton GPs within a focused range in momentum transfer that targets puzzling measurements of previous experiments. The results have provided experimental evidence that present challenges to the theoretical predictions, in particular for the electric polarizability of the proton. They have also demonstrated the potential of improving significantly our measurements for both the electric and the magnetic generalized polarizability, towards a deep understanding of the underlying dynamical mechanisms that drive these fundamental properties in the proton. The VCS-I results have also demonstrated that the systematic uncertainties of these measurements are well under control, within the level that was projected in the VCS-I proposal, as the PAC requested in the PAC-44 report. Considering the VCS-I results, along with the recommendation of PAC-44, we here submit the proposal for the VCS-II experiment. With the recent results in mind, we propose high precision measurements combined with a fine mapping as a function of $Q^2$, lower and higher in $Q^2$ with respect to the measurements of VCS-I. The proposed measurements will allow to pin down the dynamic signature of both scalar GPs with a high precision. It will also explore further the puzzling structure that has been observed in the electric GP, aiming to identify its shape that will serve as valuable input to the theory.

In the upcoming sections we present a brief introduction on the
Generalized Polarizabilities, followed by a detailed
discussion of the physics goals of the proposal, the design of the
experiment and the expected measurements.

\subsection{The Generalized Polarizabilities of the Proton}

The polarizabilities of a composite system such as the nucleon are
elementary structure constants, just as its size and shape, and can
be accessed experimentally by Compton scattering processes. In the
case of real Compton scattering (RCS), the incoming real photon
deforms the nucleon, and by measuring the energy and angular
distributions of the outgoing photon one can determine the induced
current and magnetization densities. The global strength of these
densities is characterized by the nucleon polarizabilities. In
contrast, the virtual Compton scattering (VCS) process is obtained
if the incident real photon is replaced by a virtual photon. The
virtuality of the photon allows us to map out the spatial
distribution of the polarization densities in the proton. In this case it is the
momentum of the outgoing real photon $q'$ that defines the size of
the electromagnetic perturbation while the momentum of the virtual photon $q$ sets
the scale of the observation. In analogy to the
elastic scattering, which describes the charge and magnetization
distributions, VCS gives access to the deformation of these
distributions under the influence of an electromagnetic field
perturbation as a function of the distance scale. The structure
dependent part of the process is parametrized by the Generalized
Polarizabilities (GPs), which is the generalization in four-momentum transfer space, $\alpha_E (Q^2)$ and $\beta_M(Q^2)$, of the static electric and magnetic polarizabilities obtained in RCS~\cite{gp1}.
The meaning of the generalized polarizabilities (GPs) is analogous to that of the nucleon form factors. Their Fourier transform will map out the spatial distribution density of the polarization induced by an EM field. They  represent fundamental properties of the system that probe the quark substructure of the nucleon and offer unique insight to the underlying nucleon dynamics allowing us, e.g., to study the role and the interplay of the pion cloud and quark core contributions at various length scales.
The interest on the GPs extends beyond the direct information that they provide on the dynamics of the system. They frequently enter as input parameters in a variety of scientific problems. One such example involves the hadronic two-photon exchange corrections, which are needed for a precise extraction of the proton charge radius from muonic Hydrogen spectroscopy measurements~\cite{annrev:1}. The GPs depend on the quantum numbers of the two electromagnetic transitions involved in the
Compton process and typically a multipole notation is adopted.
Initially ten independent lowest-order GPs were defined~\cite{gp2};
it was shown~\cite{gpred1,gpred2} that nucleon crossing and charge
conjugation symmetry  reduce this number to six, two scalar (S=0)
and four spin, or vector GPs (S=1). They can be defined as shown in
table~\ref{tab-dipole-gps}. The two scalar GPs, electric and
magnetic, are defined as:

\begin{eqnarray*}
\begin{array}{lll}
\alpha_E (Q^2) & = & - P^{(L1,L1)0} (Q^2) \cdot ( {e^2 \over 4 \pi} \sqrt{3 \over 2} ) \\
\beta_M  (Q^2) & = & - P^{(M1,M1)0} (Q^2) \cdot ( {e^2 \over 4 \pi} \sqrt{3 \over 8} ) \\
\end{array}
\end{eqnarray*}

\begin{table}[htbp]
\begin{center}
\caption{Notation for the six dipole GPs. In the first column the
notation uses the polarization state $\rho (\rho')$ of the initial
(final) photon, the angular momentum $L (L')$ of the transition, and
the non spin-flip $(S=0)$ or spin-flip $(S=1)$  of the nucleon. The
multipole notation in the second column uses the magnetic and
longitudinal multipoles. The six listed GPs correspond to the lowest
possible order in $q'_{cm}$, i.e. a  dipole final transition
$(l'=1)$. The third column gives the correspondence in the RCS
limit, defined by $Q^2 \to 0$ or $q_{cm} \to 0$. }
\label{tab-dipole-gps}
\begin{tabular}{|c|c|c|}
\hline \  $P^{(\rho ' L', \rho L ) S } (q_{cm})$ \ & \ $P^{(f,i)S}
(q_{cm}) \ $
& RCS limit  \\
\hline
 $P^{(01,01)0}$ & $P^{(L1,L1)0}$ &
$ - {4 \pi \over e^2} \sqrt{2 \over 3} \ \alpha_E $   \\
 $P^{(11,11)0}$ & $P^{(M1,M1)0}$ &
$ - {4 \pi \over e^2} \sqrt{8 \over 3} \ \beta_M $  \\
 $P^{(01,01)1}$ & $P^{(L1,L1)1}$ & 0 \\
 $P^{(11,11)1}$ & $P^{(M1,M1)1}$ & 0  \\
 $P^{(01,12)1}$ & $P^{(L1,M2)1}$ &
$ - {4 \pi \over e^2} {\sqrt{2} \over 3} \ \gamma_3$  \\
 $P^{(11,02)1}$ & $P^{(M1,L2)1}$ &
$ - {4 \pi \over e^2} { 2 \sqrt{2} \over 3 \sqrt{3}}  (\gamma_2 + \gamma_4)$ \\
\hline
\end{tabular}
\end{center}
\end{table}

Contrary to the form factors that describe only the ground state, the polarizabilities are sensitive to all the excited
spectrum of the nucleon. One can offer a simplistic picture of the
polarizabilities by describing the resulting effect of an
electromagnetic perturbation applied to the nucleon components. For
example, an electric field moves positive and negative charges
inside the proton in opposite directions. The induced electric
dipole moment is proportional to the electric field, and the
proportionality coefficient is the electric polarizability which
measures the rigidity of the proton. On the other hand, a magnetic
field acts differently on the quarks and the pion cloud giving rise
to two different contributions in the magnetic polarizability, a paramagnetic and a diamagnetic. Unlke the atomic polarizabilities,
which are of the size of the atomic volume, the proton electric
polarizability $\alpha_E$ \cite{gp1} is much smaller than the volume
scale of a nucleon (namely, only a few \% of its volume). The small size of
the polarizabilities underlines the extreme stiffness of the proton as
a direct consequence of the strong binding of its inner
quark and gluon constituents, offering a natural
indication of the intrinsic relativistic character of the nucleon.
In theoretical models the electric GP $\alpha_E$ is predicted
to decrease monotonically with $Q^2$. The magnetic
GP $\beta_M$ is predicted to have a smaller magnitude relative to $\alpha_E$, that can be explained by the
interplay of the competing paramagnetic and diamagnetic
contributions in the proton, which largely cancel. Furthermore, the $\beta_M$ is
predicted to go through a maximum before decreasing. This last
feature is usually explained by the dominance of diamagnetism due to
the pion cloud at long distances (small $Q^2$) and the dominance
of paramagnetism due to a quark core at short distances.

\begin{figure}[h]
\centering
\includegraphics[width=14cm]{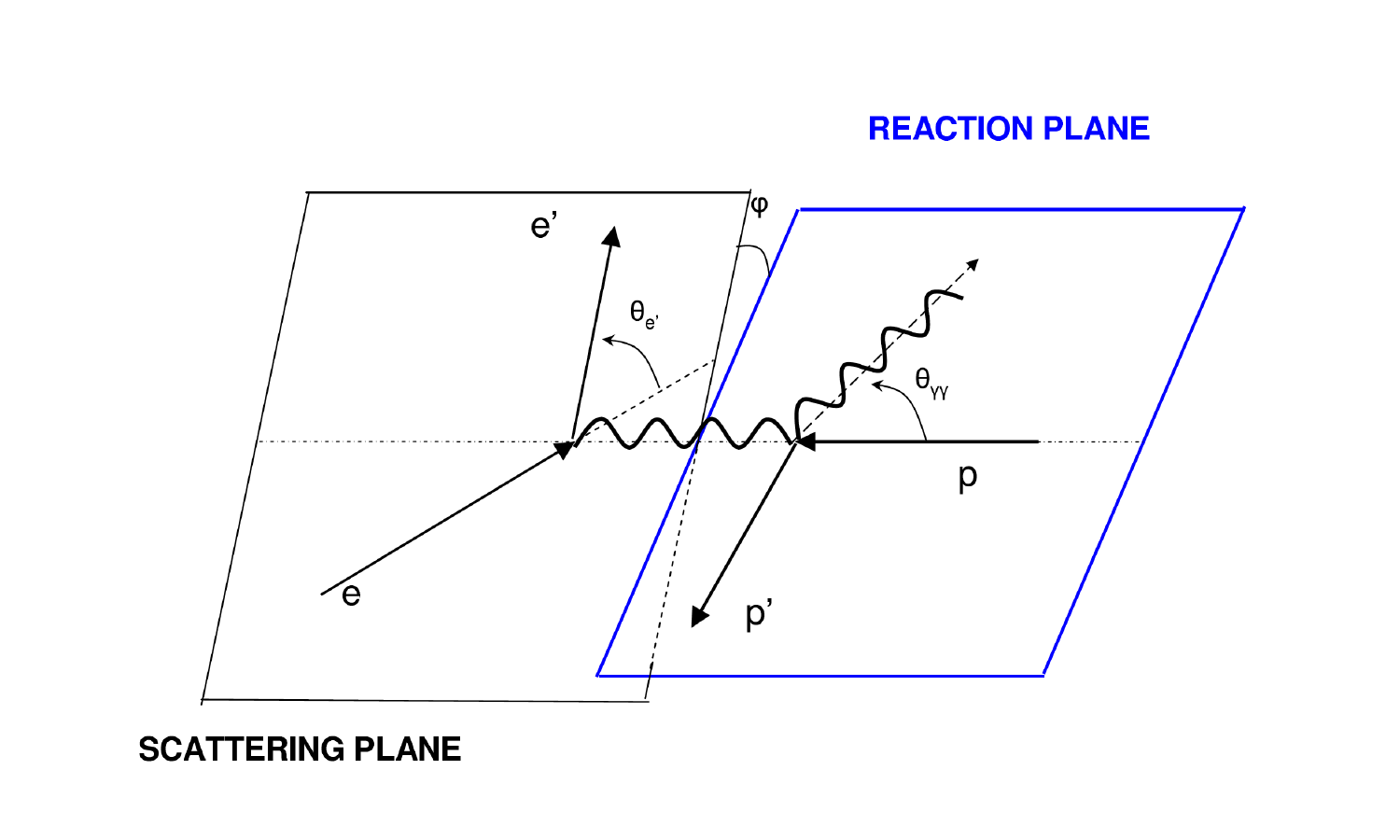}
\caption{\label{fig:vcs-reaction} The Virtual Compton Scattering
reaction}
\end{figure}

\subsection{Virtual Compton Scattering and the GPs}

\begin{figure}[t]
\centering
\includegraphics[width=10cm]{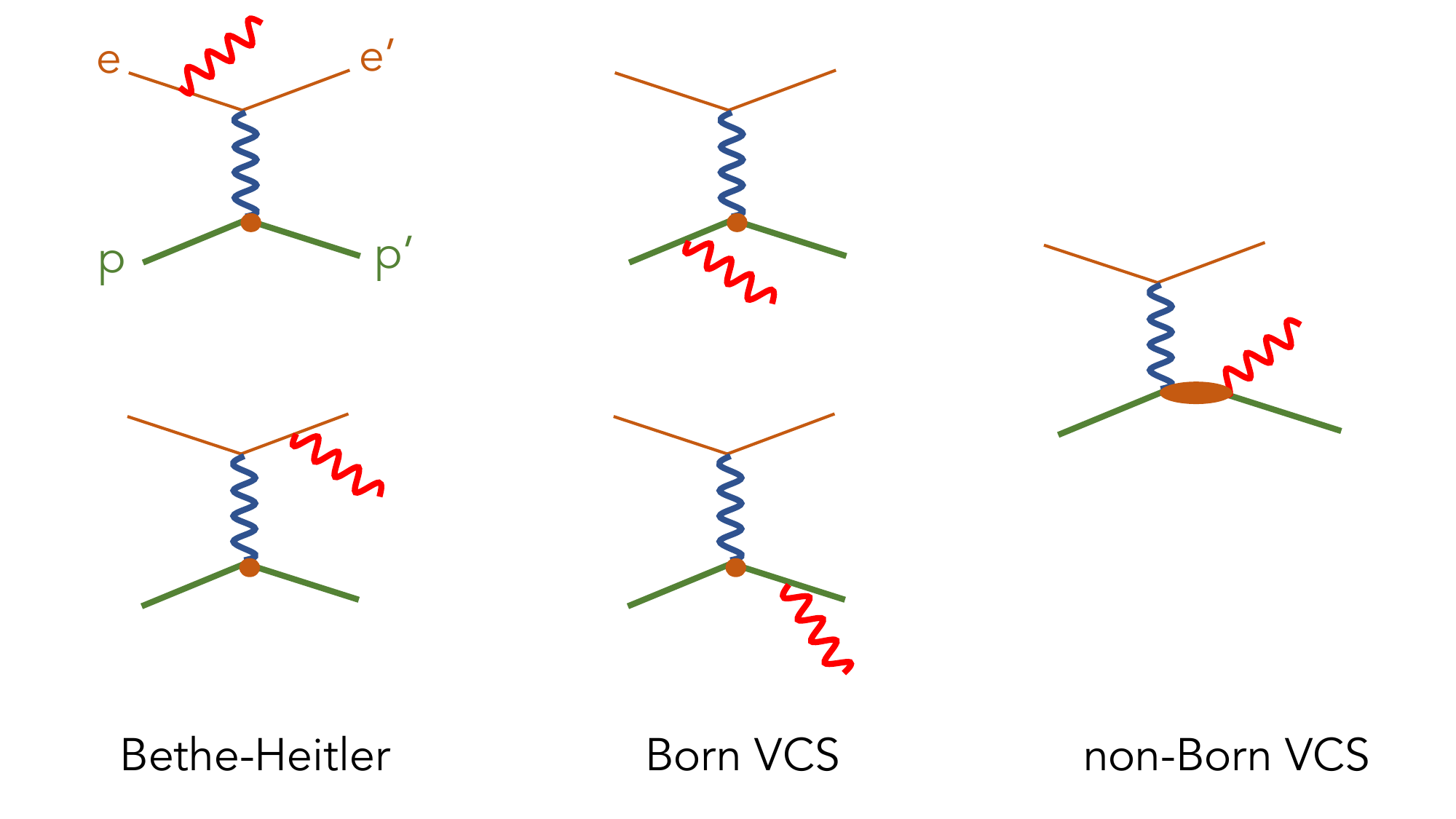}
\caption{\label{fig:vcs-diagram}Feynman diagrams of photon electroproduction, illustrating
the mechanisms contributing to ep$\rightarrow$ep$\gamma$. The small circles represent the interaction vertex of a virtual photon with a proton considered as a point-like particle, while the ellipse denotes the non-Born VCS amplitude.}
\end{figure}

The generalized polarizabilities can be explored through VCS, which is accessed experimentally
by exclusive photon electroproduction as shown in
Figure~\ref{fig:vcs-reaction}. The kinematics are defined by five
independent variables, the incoming and final electron energies, the
scattered electron angle, and the polar and azimuthal angles of the
Compton subprocess in its center-of-mass. Due to electron
scattering, one also has the Bethe-Heitler process (BH) where the
final photon is emitted by the incoming or outgoing electron. The
photon electroproduction amplitude is the coherent sum of the
Bethe-Heitler, Born and non-Born contributions as shown in
Figure~\ref{fig:vcs-diagram}. The (BH) and (Born) parts, produced
due to bremsstrahlung of the electron or proton, respectively, are
well known and are entirely calculable with the nucleon EM form
factors as inputs, while the non-Born amplitude contains the
dynamical internal structure information in terms of GPs.


The LET (Low energy theorem)~\cite{gp2} provides a path to access
these observables analytically. According to the LET, or LEX
(Low-energy EXpansion), the amplitude $T^{e p \gamma}$ is expanded
in powers of $q'_{cm}$. As a result, the (unpolarised)  $ep
\rightarrow ep \gamma$ cross section at small $q'_{cm}$ can be
written as:
\qquad \\
\begin{eqnarray}
 d^5 \sigma &=&
 d^5 \sigma ^{BH+Born}  \ + \ q'_{cm} \cdot \phi \cdot \Psi_0 \ + \ {\cal O}(q'^2_{cm}) \ \ \
\label{eq-let1}
\end{eqnarray}
\qquad \\
where $\phi$ is a phase-space factor. The notation $d^5 \sigma$
stands for $d^5 \sigma / dk'_{elab} d \Omega'_{e lab} d \Omega_{cm}$
where $k'_{elab}$ is the scattered electron momentum in the lab
frame, $d \Omega'_{e lab}$ its solid angle in the lab frame and  $d
\Omega_{\gamma cm}$  the solid angle of the outgoing photon (or
proton) in the $p$-$\gamma^*$ CM frame. The $\Psi_0$ term comes from
the interference between the Non-Born and the BH+Born amplitudes at
lowest order in $q'_{cm}$; it gives the leading polarizability
effect in the cross section. The LET approach is valid only below
the pion production threshold, i.e. as long as the Non-Born
amplitude remains real.

The $\Psi_0$ term contains three structure functions $P_{LL}$,
$P_{TT}$ and $P_{LT}$:
\qquad \\
\begin{eqnarray}
\Psi_0 &=& v_1 \cdot (P_{LL} - {\displaystyle 1 \over \displaystyle
\epsilon} P_{TT}) \ + \ v_2 \cdot  P_{LT} \label{eq-let2}
\end{eqnarray}
\qquad \\
where $\epsilon$ is the usual virtual photon polarisation parameter
and $v_1, v_2$ are kinematical coefficients depending on
$(q_{cm},\epsilon,\theta_{cm},\varphi)$. $\theta_{cm}$ and $\varphi$
are the polar and azimuthal angles of the Compton scattering process
in the CM frame of the initial proton and virtual photon
(Fig.~\ref{fig:vcs-reaction}). The full expression of $v_1,v_2 $ can
be found in ref~\cite{gp2}, as well as the expression of the
structure functions versus the GPs. For the structure functions one
has:
\qquad \\
\begin{eqnarray}
\begin{array}{lll}
P_{LL}  & = & { 4 M  \over \alpha_{em} } \cdot
G_E^p(Q^2)\cdot  \alpha_E(Q^2)   \\
\qquad \\
P_{TT}  & = & [P_{TT~spin}]     \\
\qquad \\
P_{LT}  & = & - { 2 M  \over \alpha_{em} }
 \sqrt{ {q_{cm}^2 \over Q^2} } \cdot
G_E^p(Q^2) \cdot \beta_M(Q^2) +  [P_{LT~spin}]  \label{eq-sf12}
\end{array}
\end{eqnarray}
\qquad \\
where $\alpha_{em}$ is the fine structure constant and the terms in
brackets are the spin part of the structure functions:

\begin{eqnarray}
\begin{array}{lll}
P_{TT spin}  & = &  -3 G_M^p(Q^2) \, {q_{cm}^2 \over {\tilde q^0} }
\cdot
(P^{(M1,M1)1}(Q^2) \\
\ & \ &  - {\sqrt{2}} \, {\tilde q^0} \cdot P^{(L1,M2)1}(Q^2)  \\
\qquad \\
P_{LT spin}  & = & {3 \over 2} \, {  q_{cm} \sqrt{ Q^2} \over \tilde
q^0 }  G_M^p(Q^2) \cdot P^{(L1,L1)1}(Q^2)  \label{eq-sfspin}
\end{array}
\end{eqnarray}
\qquad \\
where $\tilde q^0$ is the CM energy of the virtual photon in the
limit $q'_{cm} \to 0$. One can note that $P_{LL}$ is proportional to
the electric GP, and the scalar part of $P_{LT}$ is proportional to
the magnetic GP. Using this LET approach one cannot extract all six
dipole GPs separately from an unpolarised experiment since only
three independent structure functions appear and can be extracted
assuming the validity of the truncation to ${\cal O}(q'^2_{cm})$.
Furthermore in order to isolate the scalar part in these structure
functions a model input is also required.


However, since the sensitivity of the VCS cross sections to the GPs
grows with the photon energy it is advantageous to go to higher
photon energies. Above the pion threshold the VCS amplitude becomes
complex. While $T^{BH}$ and $T^{Born}$ remain real, the amplitude
$T^{Non-Born}$ acquires an imaginary part, due to the coupling to
the $\pi$N channel. The relatively small effect of GPs below the
pion threshold, which is contained in $d\sigma_{Non-Born}$, becomes
more important in the region above the pion threshold and up to the
$\Delta(1232)$ resonance, where the LET does not hold. In this case
a Dispersion Relations (DR) formalism is prerequisite to extract the
polarizabilities in the energy region above pion threshold where the
observables are generally more sensitive to the GPs.


The Dispersion Relations (DR) formalism developed by B.Pasquini et
al. \cite{gp12,gp13} for RCS and VCS allows the extraction of
structure functions and GPs from photon electroproduction
experiments. The calculation provides a rigorous treatment of the
higher-order terms in the VCS amplitude, up to the $N \pi \pi$
threshold, by including resonances in the $\pi N$ channel. The
Compton tensor is parameterised through twelve invariant amplitudes
$F_{i} (i=1,12)$. The GPs are expressed in terms of the non-Born
part  $F_i^{NB}$ of these amplitudes at the point $t=-Q^2,
\nu=(s-u)/4 M  =0$, where $s,t,u$ are the Mandelstam variables of
the Compton scattering.
%
%
%
%
All of the $F_i^{NB}$ amplitudes, with the exception of two, fulfill
unsubtracted dispersion relations. These $s$-channel dispersive
integrals are calculated through unitarity. They are limited to the
$\pi N$ intermediate states, which are considered to be the dominant
contribution for describing VCS up to the $\Delta(1232)$ resonance
region.
The calculation uses pion photo- and electroproduction multipoles
\cite{maid} in which both resonant and non-resonant production
mechanisms are included. The amplitudes $F_1$ and $F_5$ have an
unconstrained part beyond the $\pi N$ dispersive integral. Such a
remainder is also considered for $F_2$. For $F_5$ this asymptotic
contribution is dominated by the t-channel $\pi^0$ exchange, and
with this input all four spin GPs are fixed.
For $F_1$ and $F_2$, an important feature is that in the limit
$(t=-Q^2, \nu=0)$ their non-Born part is proportional to the GPs
$\beta_M$ and $( \alpha_E + \beta_M )$,  respectively. The remainder
of $F _{1,2}^{NB}$ is estimated by an energy-independent function,
noted $\Delta \beta$ and $\Delta (\alpha + \beta)$ respectively.
This term parameterises the asymptotic contribution and/or
dispersive contributions beyond $\pi N$. For the magnetic GP one
gets:
\qquad \\
\begin{eqnarray}
\begin{array}{lll}
\beta_M (Q^2) & = & \beta^{\pi N}(Q^2) +  \Delta \beta \\
\qquad \\
 \Delta \beta  & = & { \displaystyle [ \beta^{exp}   -
\beta^{\pi N} ]_{Q^2=0} \over \displaystyle ( 1 + Q^2/
\Lambda^2_\beta )^2 } \ . \label{eq-dr-beta-0}
\end{array}
\end{eqnarray}
\qquad \\
The sum $( \alpha_E + \beta_M )$ follows a similar decomposition,
and thus the electric GP too:
\qquad \\
\begin{eqnarray}
\begin{array}{lll}
\alpha_E (Q^2) & =  &  \alpha^{\pi N}(Q^2) +  \Delta \alpha \\
\qquad \\
\Delta \alpha & = & { \displaystyle [ \alpha^{exp}   - \alpha^{\pi
N} ]_{Q^2=0} \over \displaystyle ( 1 + Q^2/ \Lambda^2_\alpha )^2 } \
. \label{eq-dr-alpha-0}
\end{array}
\end{eqnarray}
\qquad \\


The two scalar GPs are not fixed by the model, and their
unconstrained part is parametrised by a dipole form, as given by
eqs.(\ref{eq-dr-beta-0},\ref{eq-dr-alpha-0}). This dipole form is
arbitrary while the mass parameters $\Lambda_\alpha$ and
$\Lambda_\beta$ only play the role of intermediate quantities in
order to extract VCS observables. In the DR calculation
$\Lambda_\alpha$ and $\Lambda_\beta$ are treated as free parameters,
which can furthermore vary with $Q^2$. Their value can be adjusted
by a fit to the experimental cross section, separately at each
$Q^2$. Then the calculation is fully constrained and provides all
VCS observables, the scalar GPs as well as the structure functions,
at this $Q^2$.


\begin{figure}[t]
\centering
\includegraphics[width=14.5cm]{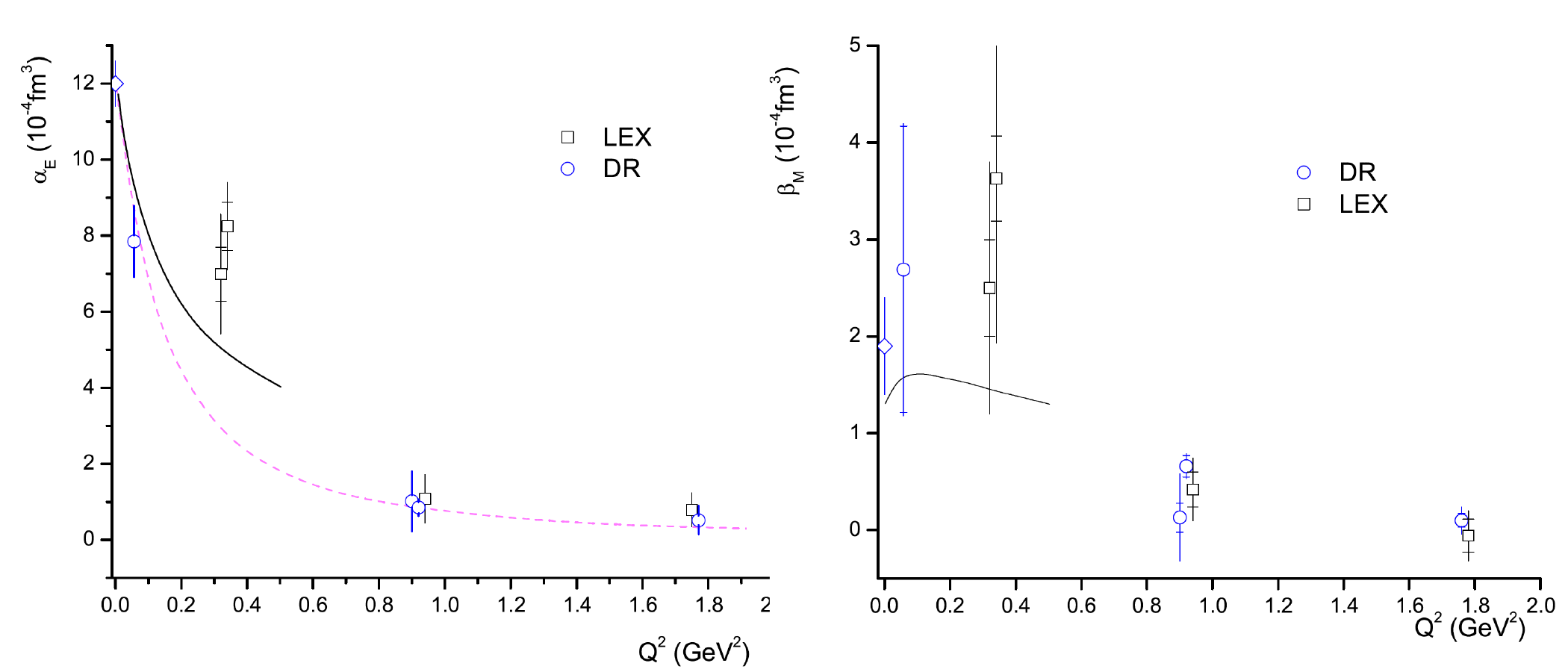}
\caption{\label{fig:aEbM} World
data from the early experiments \cite{gp3,gp4,gp5,jlabgp,gp6,gp19} on the electric GP $\alpha_E$
and the magnetic GP $\beta_M$. Open
circles correspond GPs extracted through DR while the open boxes
through LEX. The solid curve corresponds to HBChPT~\cite{gp8,gp9}.
The dipole fall off of $\alpha_E$ (dashed line) from the DR
calculation~\cite{gp12,gp13} is able to describe all world data
except the MAMI measurements.}
\end{figure}

\begin{figure}[h]
\centering
\includegraphics[width=16cm]{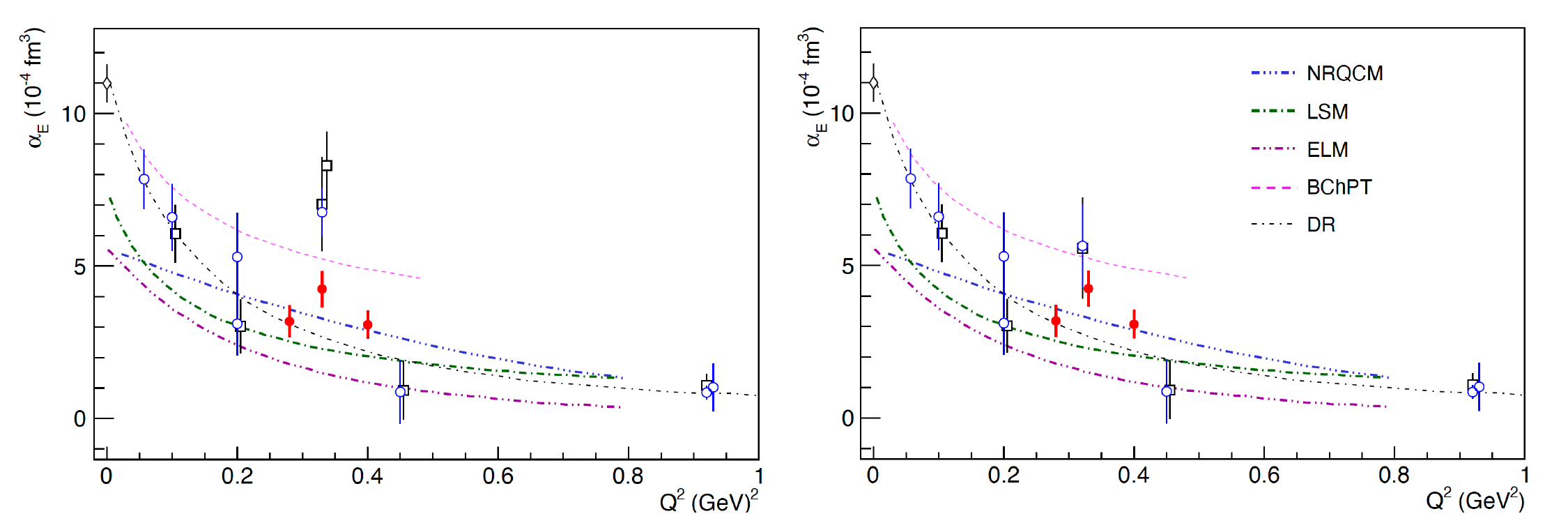}
\caption{\label{fig:aE-world} Left panel: The world
data for $\alpha_E$. The recent results from MAMI~\cite{gp21,gp22,gp20} are shown as open symbols and those of the JLab VCS-I experiment~\cite{gp23} as filled red circles. The results from the early experiments (already referenced in Fig.~\ref{fig:aEbM}) are also shown with open symbols. Right panel: Same as in the left panel, but with the re-analysis of the MAMI data at $Q^2=0.33~GeV^2$~\cite{ranal1,ranal2}. The theoretical predictions of~\cite{gp14,gp15,gp15b,gp16,gp17,gp18,kim,lensky} are also shown.}
\end{figure}

\subsection{The experimental and theoretical landscape of the GPs}

A group of early VCS experiments, that was performed at MAMI
\cite{gp3,gp4}, JLab \cite{gp5,jlabgp} and Bates \cite{gp6,gp19} nearly two decades ago,
shaped a first understanding of the proton electric and
magnetic GPs. They involved measurements below and above the pion threshold, that were analyzed both within the LEX and the DR approach. The measurements illustrated the good agreement in the extraction of the GPs following the two different methods, as well as the consistency in the GPs extraction from measurements that were conducted below and above the pion threshold, and up to
the first resonance region \cite{gp5,jlabgp}. The measurements
have furthermore highlighted the enhanced sensitivity to the GPs
as one measures above the pion production threshold. These early measurements, as illustrated in Fig.~\ref{fig:aEbM}, 
presented experimental evidence that contradict the naive Ansatz of a single-dipole
fall-off for $\alpha_E(Q^2)$, pointing out
to an enhancement at low $Q^2$ evidenced by the MAMI measurements
\cite{gp3,gp4}. Here, two independent experiments~\cite{gp3,gp4} were
able to confirm this unexpected structure for $\alpha_E$. For the magnetic GP, a first experimental description with relatively large uncertainties was provided by these experiments, highlighting the challenges in extracting the magnetic polarizability signal.

\begin{figure}[h]
\centering
\includegraphics[width=8cm]{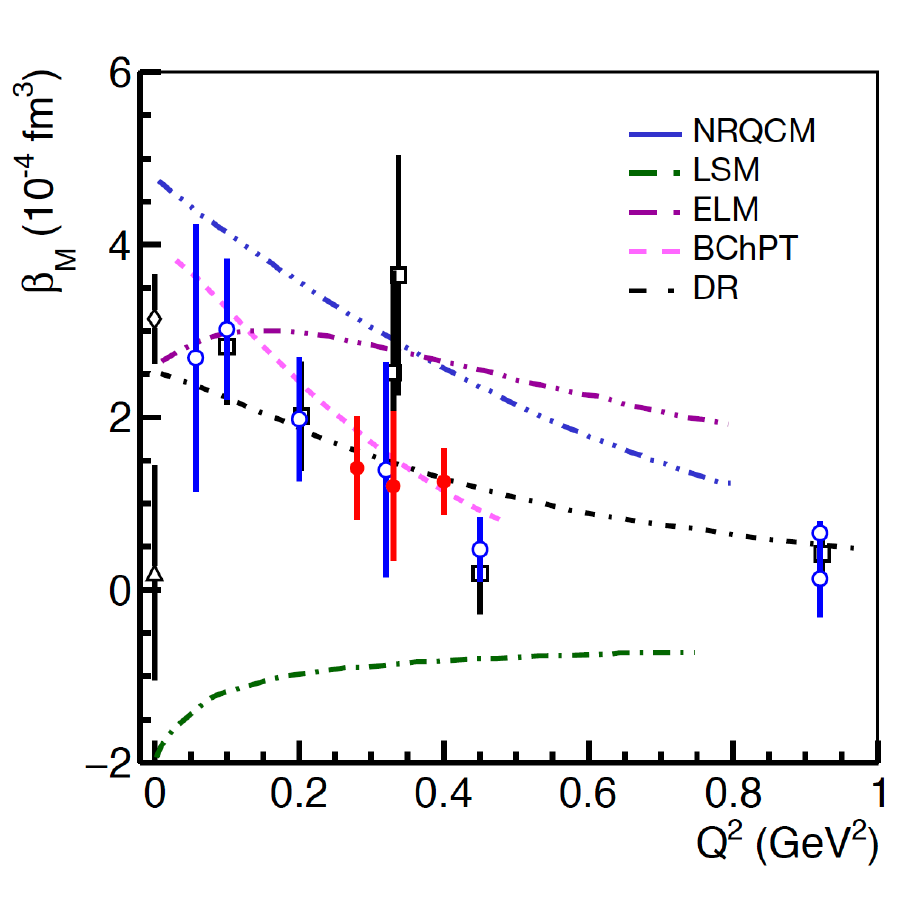}
\caption{\label{fig:betam} The magnetic generalized polarizability. The references are the same as in Fig.~\ref{fig:aE-world}.}
\end{figure}

\begin{figure}[t]
\centering
\includegraphics[width=16cm]{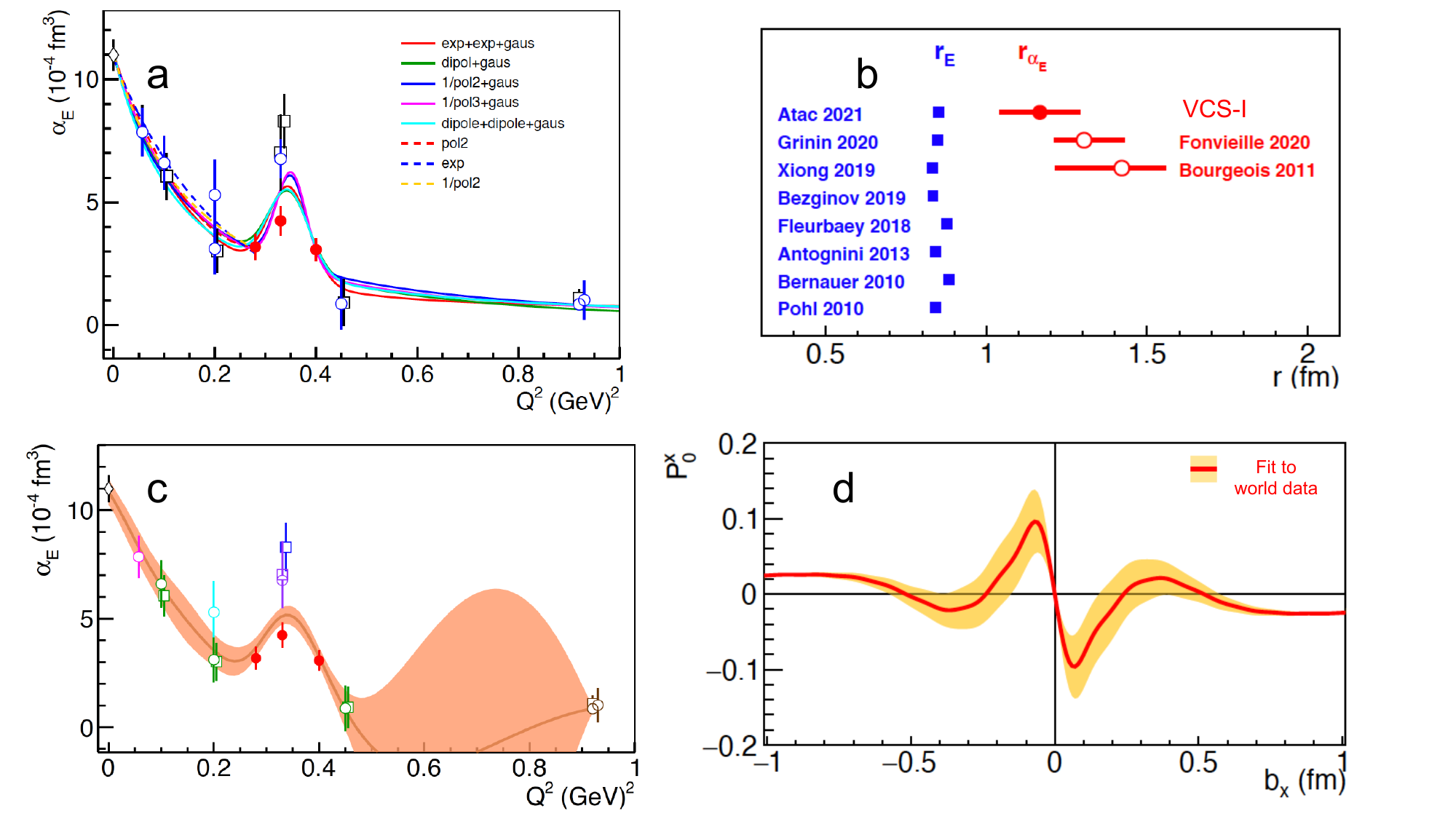}
\caption{\label{fig:alphae} {\bf a)} Mean square electric polarizability radius fits using combinations of different functional forms. The fits denoted with solid lines were performed over the full $Q^2$-range of the world-data. The functional forms denoted with dashed lines were performed in the low-$Q^2$ range, namely in $Q^2$=[0,0.28]~GeV$^2$.{\bf b)} The proton electric polarizability radius $r_{\alpha_E} \equiv \sqrt{\langle r^2_{\alpha_E} \rangle}$  is shown in red. The measurements of the proton charge radius are shown for comparison in blue color. {\bf c)} The $Q^2$-dependence of the electric GP as derived from the experimental measurements using the GPR technique~\cite{gpr}, a data-driven method that assumes no direct underlying functional form.  {\bf d)} Induced polarization in the proton when submitted to an EM field as a function of the transverse position with photon polarization along the x axis for $b_y=0$. The x-y defines the transverse plane, with the z axis being the direction of the fast moving protons. }
\end{figure}

A recent generation of experiments provided an improved and more extended study of the proton GPs. The experiments were conducted at MAMI~\cite{gp21,gp22,gp20} and at JLab~\cite{gp23} and their results are shown in Figs.~\ref{fig:aE-world} and~\ref{fig:betam}. The reported measurements provided evidence of a local structure (e.g. an enhancement or plateau) of $\alpha_E$, at the same $Q^2$ as previously reported in~\cite{gp3,gp4}, but with a smaller magnitude than what was originally suggested. In parallel, the data analysis of the early MAMI experiment~\cite{gp3,gp4} was revisited, accounting for refinements in the analysis procedure that were developed recently and were considered in the analysis of the latest MAMI measurements~\cite{gp21,gp22}. The re-analysis of these data (as presented in~\cite{ranal1,ranal2}, currently unpublished) reduces slightly the extracted value for $\alpha_E$ and brings it in agreement with the JLab VCS-I results~\cite{gp23}, as shown in Fig.~\ref{fig:aE-world}~(right panel). A local non-trivial structure in $\alpha_E$, as suggested by both the JLab and the MAMI measurements, presents a striking contradiction to our theoretical understanding. The mesonic cloud effects may present a potential candidate for the processes involved in this effect, but the presence of a dynamical mechanism that is not accounted for in the theory can't be excluded. The signature of this effect has so far been explored~\cite{gp23} with phenomenological fits as well as with methods that do not assume any direct underlying functional form~\cite{gpr}, as shown in Fig.~\ref{fig:alphae}(a) and Fig.~\ref{fig:alphae}(c) respectively. In light of the recent results, the need for new experimental measurements becomes evident. They will allow to increase the statistical significance in the observation of the local enhancement in $\alpha_E$, that deviates from the theoretically predicted monotonic $Q^2$~dependence, currently at the 3$\sigma$ level. The shape and the dynamical signature of this structure needs to be clearly mapped with  high precision measurements, so that it can serve as an input for the theory towards explaining the effect. The motivation for new measurements extends further, to the study of the magnetic GP. Here, the magnitude of the signal is smaller compared to $\alpha_E$, as shown in Fig.~\ref{fig:betam}. This renders the $\beta_M$ more sensitive to the experimental errors and challenges our access in the physics of interest. Reported discrepancies~\cite{bstat1,bstat2} in the measurement of the static ($Q^2=0$) magnetic polarizability, as shown in Fig.~\ref{fig:betam}, highlight the need to improve these measurements, particularly at low momentum transfers. Recent experiments have illustrated our ability to measure $\beta_M$ with high precision. Extending these measurements will provide a key to understand the processes manifesting in the interplay between diamagnetism and paramagnetism in the proton, and represents an excellent opportunity to gain a deep insight to the structure and the dynamics of the nucleon.

\begin{figure}[t]
\centering
\includegraphics[width=17cm]{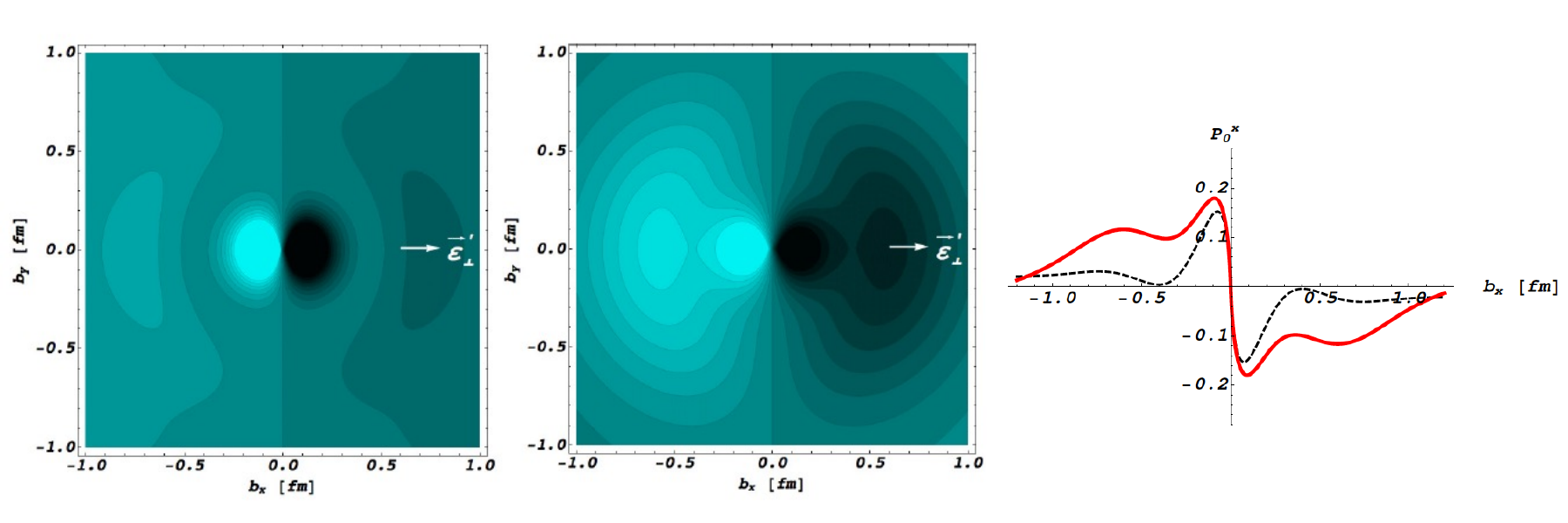}
\caption{\label{fig:pol-dens}Figure from~\cite{spa2}. Induced
polarization, $P_0^x$, in a proton of definite light-cone helicity,
when submitted to an e.m. field with photon polarization along the
$x$-axis, as indicated. Left (center) panel is for GP~I (GP
II), see text. The light (dark) regions correspond to the largest
(smallest) values. The right panel compares $P_0^x$ along $b_y =
0$~: dotted curve is for GP I; solid curve is for GP II. }
\end{figure}

In analogy to the relation that connects the proton electric form factor to the proton charge radius, the generalized polarizabilities allow access to the electromagnetic polarizability radii of the proton. The mean square electric polarizability radius of the proton $\langle r^2_{\alpha_E} \rangle$ is related to the slope of the electric GP at $Q^2=0$ by

\begin{equation}\label{radius} 
\langle r^2_{\alpha_E} \rangle = \frac{-6}{\alpha_E(0)} \cdot \frac{d}{dQ^2} \alpha_E(Q^2) {\bigg \vert}_{Q^2=0}.
\end{equation}

In a procedure that is equivalent to that for the extraction of the proton charge radius, fits to the world-data~\cite{gp23} of the electric generalized polarizabilities, as shown in Fig.~\ref{fig:alphae}(a), lead to a value for the mean square electric polarizability radius $\langle r^2_{\alpha_E}\rangle=1.36 \pm 0.29~fm^2$. This value is considerably larger compared to the mean square charge radius of the proton, $\langle r^2_{E} \rangle \sim0.7~fm^2$ (see Fig.~\ref{fig:alphae}(b)). The dominant contribution to this effect is expected to arise from the deformation of the mesonic cloud in the proton under the influence of an external EM field.
Similarly, the mean square magnetic polarizability radius has been extracted from the magnetic polarizability measurements $\langle r^2_{\beta_M}\rangle=0.63 \pm 0.31~fm^2$. A new set of measurements for the proton electromagnetic GPs will allow to improve further the precision in the extraction of the proton polarizability radii.

Significant theoretical progress has been achieved in recent years towards our understanding of the generalized polarizabilities. The GPs have been calculated following a variety of approaches, as shown in Fig.~\ref{fig:aE-world}.
A common feature is that none of the theoretical calculations is able to describe a non-trivial structure of $\alpha_E$, as they all predict a smooth $Q^2$ fall-off. In heavy baryon chiral perturbation theory
(HBChPT) the polarizabilities are pure one-loop effects to leading
order in the chiral expansion~\cite{chpt}, emphasizing the role of
the pion cloud; the scalar GPs have been calculated to order
$p^3$~\cite{gp7,gp8,gp9}, while the spin GPs have been calculated to
order $p^4$~\cite{gp10,gp11}. The first nucleon resonance $\Delta
(1232)$ is taken into account either by local counterterms
(ChPT,~\cite{chpt}) or as an explicit degree of freedom (small scale
expansion SSE of~\cite{gp9}). In non-relativistic quark constituent
models~\cite{gp2,gp14,gp15,gp15b} the GPs involve  the summed
contribution of all nucleon resonances but do not embody a direct
pionic effect. The calculation of the linear-$\sigma$
model~\cite{gp16,gp17} involves all fundamental symmetries but does
not include the $\Delta$ resonance, while the effective lagrangian
model~\cite{gp18} includes resonances and the pion cloud in a more
phenomenological way. A calculation of the electric GP was made in
the Skyrme model~\cite{kim}. Recent calculations of the generalized polarizabilities have been performed in baryon chiral perturbation theory~\cite{lensky}.
Lattice calculations are for the moment
limited to polarizabilities in RCS~\cite{det} but significant
progress on that front is expected in the near future. Future experimental measurements, such as those presented in this proposal, will provide high-precision benchmark data for these calculations, offering valuable input and guidance to the theory.

The formalism to extract light-front quark
charge densities from nucleon form factor data has been extended in recent years to the
deformation of the quark charge densities when applying an
external electric field ~\cite{spa1,spa2}. This in-turn allows for
the concept of GPs to be used to describe the spatial deformation of
the charge and magnetization densities. The correlation of the GPs to the induced
polarization in a proton when submitted to an e.m. field is
illustrated in the theoretical calculations shown in Fig.~\ref{fig:pol-dens}, where two parametrizations, GP-I (dipole fall-off)
and GP-II (a parametrization of a dipole+gaussian), have been considered for $\alpha_E$~\cite{param}. An
$\alpha_E$ enhancement at intermediate $Q^2$, as opposed to a pure dipole fall-off,
tends to extend the spatial distribution of the induced
polarization to larger transverse distances. An experimental extraction of the induced polarization in the proton from a fit to the world data~\cite{gp23} is shown in Fig.~\ref{fig:alphae}(d). Upcoming experiments will allow to improve further the precision of this extraction, offering a detailed spatial representation of the induced polarization in the proton.

\section{The Experiment}

The proposed experiment aims to provide high precision measurements of $\alpha_E$ and $\beta_M$ in the
region of $Q^2=0.05~(GeV/c)^2$ to $Q^2=0.50~(GeV/c)^2$. The proposed measurements will allow to pin down the dynamic signature of the two scalar GPs with a high precision, providing access to the underlying reaction mechanisms in the proton. They will further explore the puzzling structure that has been observed in the electric GP, aiming to identify the shape of this structure that will serve as valuable input to the theory. More specifically, the selection of the proposed kinematics has been done considering recent measurements and findings for the proton GPs, along with the demonstrated potential to improve upon the precision of the world data with measurements at JLab, towards the following goals:

i) Provide high precision measurements combined with a fine mapping as a function of $Q^2$, lower and higher in $Q^2$ with respect to the kinematics that are of particular interest for $\alpha_E$. This is vital, since it will allow to explore the dynamical signature of  $\alpha_E$ through a set of measurements that are all uniform in regard to their systematic uncertainties. This will in-turn provide further clarity in the study of the suggested structure for the $\alpha_E$ and in identifying it's shape with precision.

ii) Provide measurements within targeted kinematics where the sensitivity to the polarizabilities is appreciably changing. For the particular case of  $\alpha_E$, 
complementary measurements at the same $Q^2$ will be conducted within kinematics where the suggested structure in the polarizability emerges in an anti-diametric way in the VCS cross section. These measurements will allow to fully de-couple the observation of a non-trivial structure in the polarizability from the influence of experimental uncertainties that are of systematic nature. 

iii) Improve significantly the precision in the measurement of the $\beta_M$. The world data for $\beta_M$ are characterized by relatively large uncertainties and they do not provide sufficient information to decode the interplay of the paramagnetic and diamagnetic mechanisms in the proton, that are particularly profound in the low~$Q^2$ region. Recent measurements at $Q^2=0$~\cite{bstat1,bstat2} reported discrepancies in the measurements of the magnetic GP. They  have added to the world-data tension and have further challenged our understanding of $\beta_M$. The results from VCS-I~\cite{gp23} have illustrated that we can improve tremendously upon our knowledge of the magnetic GP through a follow-up series of measurements, as proposed in this work.

\subsection{The experimental setup}

\begin{figure}[h]
\centering
\includegraphics[width=18.0 cm]{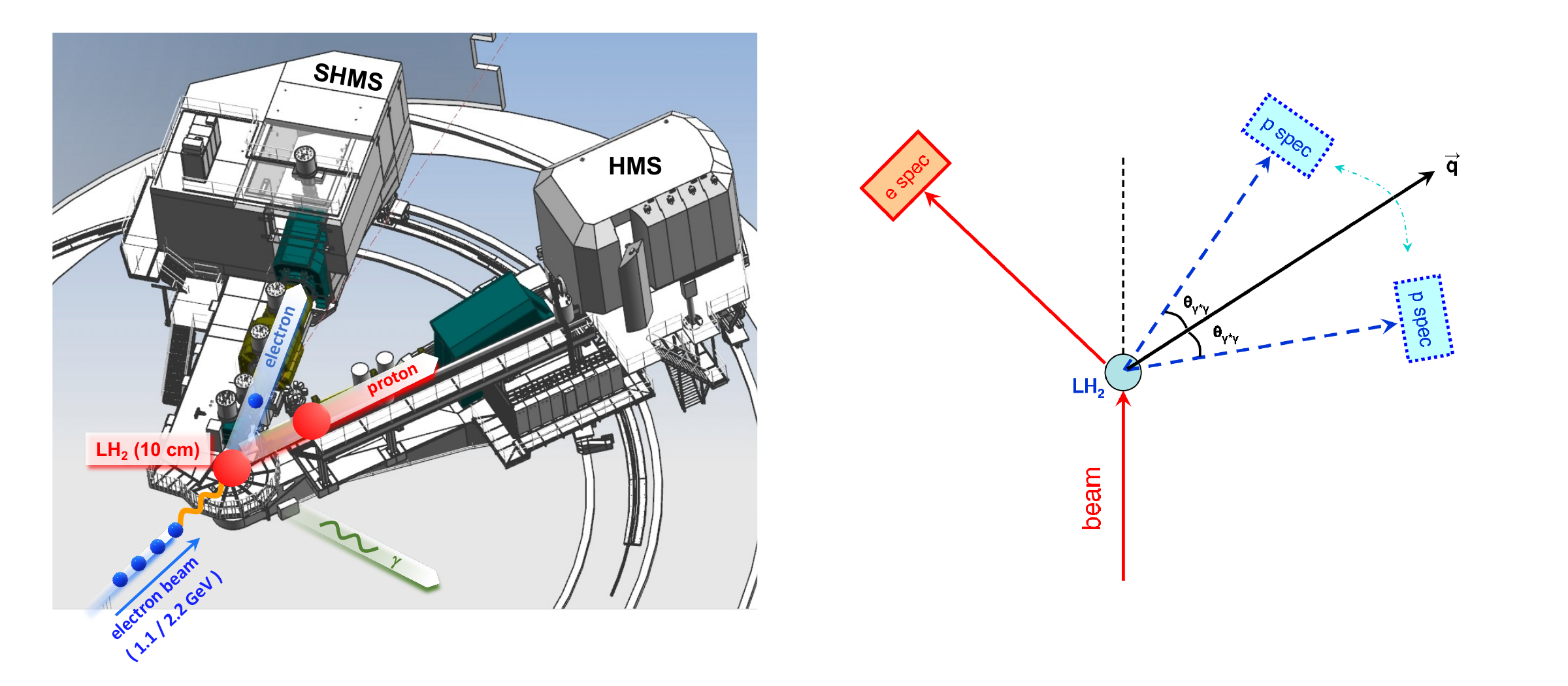}
\caption{Left: An illustration of the proposed experimental setup in Hall~C.  Right: Positioning the proton spectrometer symmetrically with respect to the momentum transfer direction allows to measure the azimuthal asymmetry of the VCS cross section, where part of the systematic uncertainties get suppressed. By keeping both magnet settings unchanged between the two asymmetry measurements, the spectrometer momenta can be precisely determined through the cross-calibration of the two reconstructed missing mass spectra.   
\label{hallc_fig}}
\end{figure}

The experiment will utilize standard Hall C equipment to provide
measurements of the Virtual Compton Scattering. A schematic representation of the experimental setup is presented in
Fig.~\ref{hallc_fig}. The SHMS and HMS
spectrometers~\cite{shms,hms} will be used to detect electrons and
protons in coincidence respectively, while the reconstructed missing
mass will provide the identification of the photon. An electron beam
of $E_\circ=1.1$ and $2.2~GeV$ with current up to $I=75\mu$A, along with a 10~cm long liquid
hydrogen target will be required for this measurement. The GPs will be measured within the range of
$Q^2=0.05~(GeV/c)^2$ to $Q^2=0.50~(GeV/c)^2$.

\begin{figure}
\begin{center}
\includegraphics[width = 18.5cm]{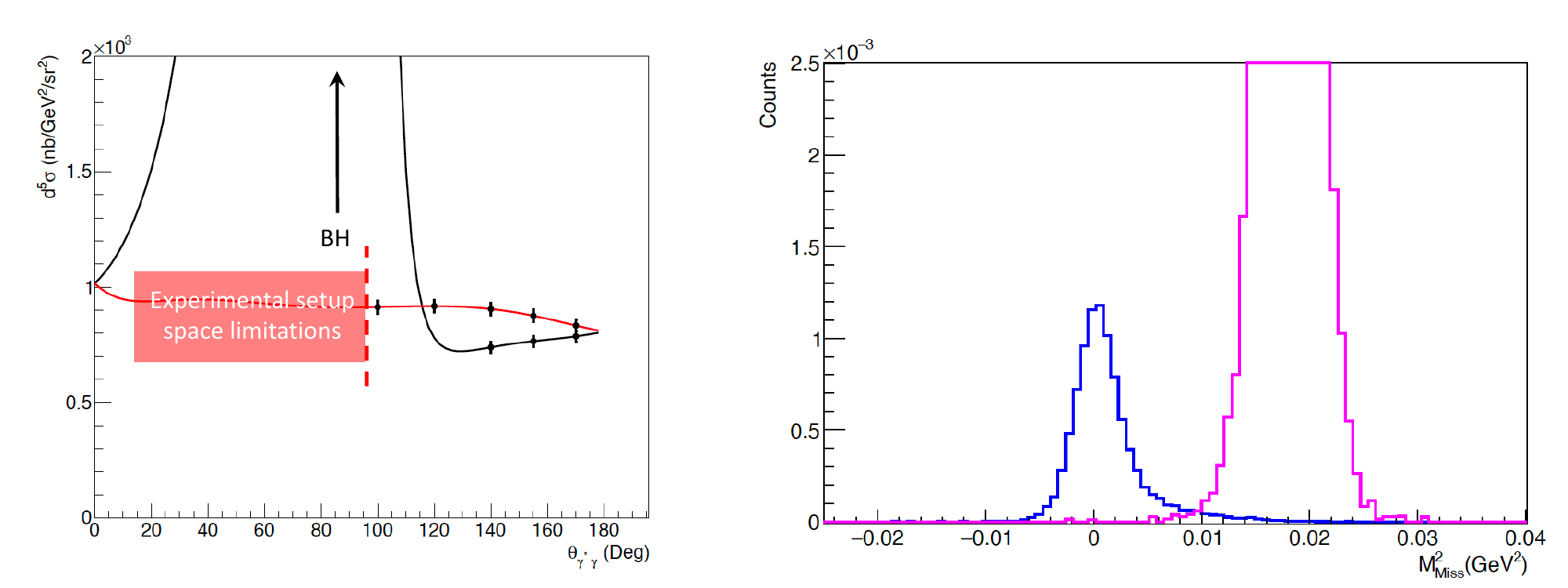}
\end{center}
\vspace{-0.35cm} \vspace{-0.35cm} \caption{Left panel: Projected
cross sections at $Q^2=0.35~(GeV/c)^2$. The black and red curves
correspond to cross sections at $\phi_{\gamma^*\gamma}=0^\circ$ and
$180^\circ$. The constraints imposed by the BH and by the space limitations of the experimental setup are marked in the figure. Right
panel: The reconstructed missing mass spectrum offers a clean separation between the undetected photons and the pions.} \label{fig:q43}
\end{figure}

A beam energy of $E_\circ=1.1~GeV$ is required only for the lowest $Q^2$ setting, at $Q^2=0.05~(GeV/c)^2$. All the other measurements request a beam of $E_\circ=2.2~GeV$. The value for the beam energy has been chosen so that it can simultaneously accommodate the maximum beam energy of the accelerator in another Hall. Nevertheless, its exact value can be flexibly adjusted as needed, with a minimal impact in the extraction of the GPs. The choice of the kinematic settings has been driven by the sensitivity of the Compton scattering observables to the polarizabilities, in tandem with physical constraints of the experimental setup. The measurements will target kinematics above the pion threshold, that offer enhanced sensitivity to the GPs. They will span a wide range of energies and photon angles, typically above $\theta_{\gamma^*\gamma}\approx$~100$^\circ$ in order to satisfy the physical restrictions in the Hall and to avoid the kinematic range where the BH process dominates the cross section (e.g. see
Fig.~\ref{fig:q43}) and suppresses the sensitivity to the GPs. Measurements of the in-plane azimuthal asymmetry of the VCS cross section with respect
to the momentum transfer direction will be conducted, for fixed $\theta_{\gamma^*\gamma}$ at $\phi_{\gamma^*\gamma}=0^\circ$ and
$\phi_{\gamma^*\gamma}=180^\circ$
\qquad \\
$A_{(\phi_{\gamma^*\gamma}=0,\pi)} =
\frac{\sigma_{\phi_{\gamma^*\gamma}=0} - \sigma_{\phi_{\gamma^*
\gamma}=180}}  {\sigma_{\phi_{\gamma^*\gamma}=0} +
\sigma_{\phi_{\gamma^*\gamma}=180}}$
\qquad \\
One benefit emerging from the asymmetry measurements is that part of the systematic uncertainties are suppressed in the ratio, and thus the sensitivity in the extraction of the polarizabilities is enhanced. Furthermore, the asymmetry measurements allow to precisely determine the true momentum settings of the two spectrometers 
based on the cross-calibration of the missing mass, since the momentum and position of the electron spectrometer remain the same between the two settings while the momentum setting for the proton spectrometer also remains unchanged. This allows to correct for small deviations between the set and the true spectrometer momentum settings and offers a tighter control of the systematic uncertainties.

\begin{figure}[t]
\begin{center}
\includegraphics[width = 9cm]{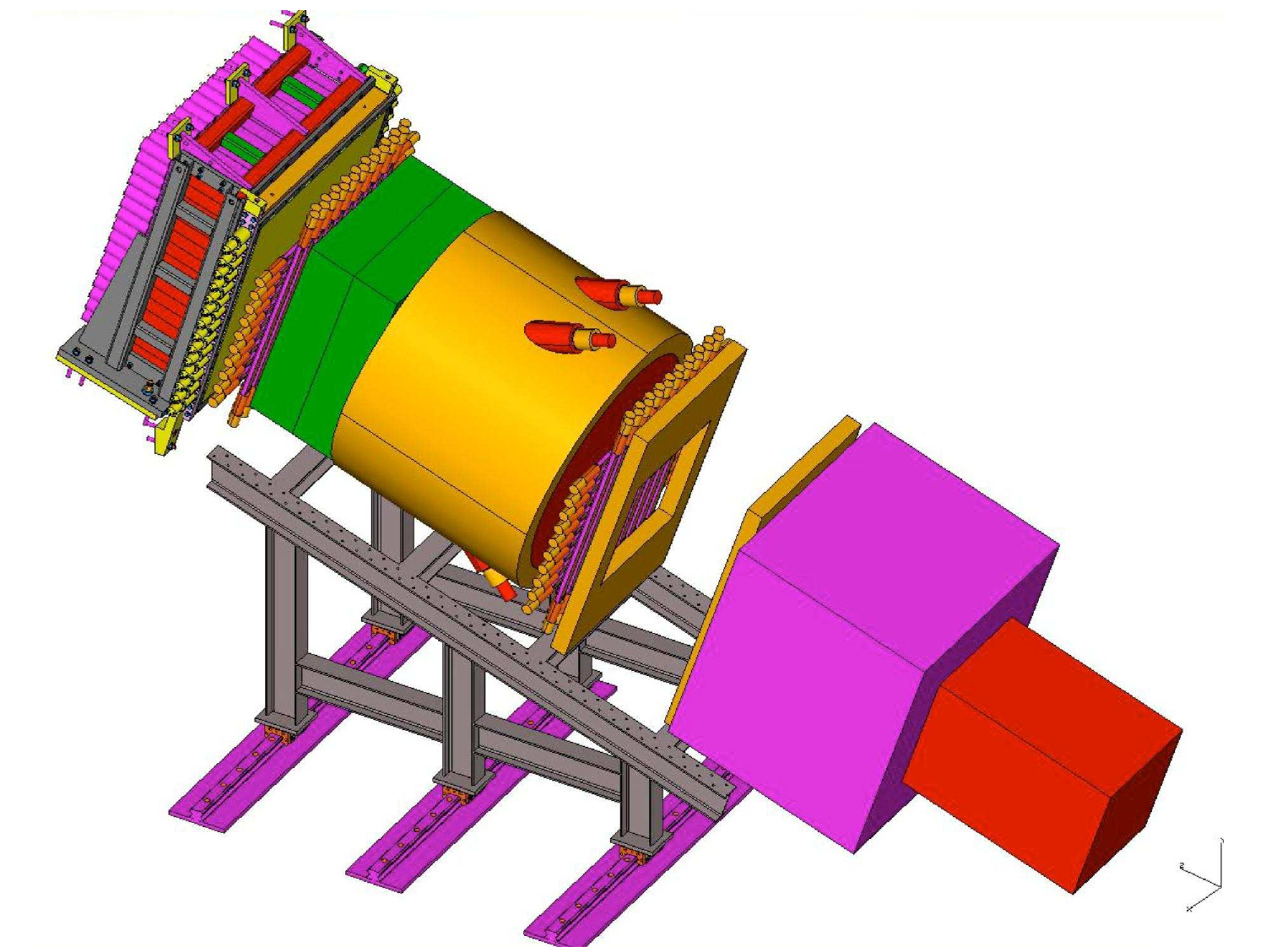}
\end{center}
\caption{\label{fig:shms} The SHMS detector stack.}
\end{figure}

\begin{center}
\begin{table} \large
\begin{tabular}{|c|c|c|}
\hline
  Kinematic   & Kinematic          &  HMS singles rates\\
  Group   & Setting              &  (kHz)\\
\hline
           &   Kin I               &  43      \\
           &   Kin II              &  53      \\
  GI       &   Kin IIIa            &  119     \\
           &   Kin IIIb            &   65     \\
           &   Kin IVa             &  128     \\
           &   Kin IVb             &  80      \\

\hline           
           &   Kin I               & 159      \\
           &   Kin IIa             &  21      \\
    GII    &   Kin IIb             & 155      \\
           &   Kin IIIa            &  28      \\
           &   Kin IIIb            &  122     \\
           &   Kin IVa             &  42      \\
           &   Kin IVb             &   82     \\
\hline
           &   Kin I               &  347     \\
           &   Kin IIa             &  27      \\
  GIII     &   Kin IIb             &  330     \\
           &   Kin IIIa            &  47       \\
           &   Kin IIIb            &  214      \\
           &   Kin IVa             &  77       \\
           &   Kin IVb             &  129      \\
\hline
           &   Kin I               &  476      \\
           &   Kin II              &  497      \\
GIV        &   Kin IIIa            &  453      \\
           &   Kin IIIb            &  64       \\
           &   Kin IVa             &  313      \\
           &   Kin IVb             &  89       \\
           &   Kin Va              &  212      \\
           &   Kin Vb              &  127      \\        
\hline
           &   Kin I               &  483      \\
           &   Kin II              &  502      \\
  GV       &   Kin IIIa            &  444      \\
           &   Kin IIIb            &  51       \\
           &   Kin IVa             &  295      \\
           &   Kin IVb             &  72       \\
           &   Kin Va              &  192      \\
           &   Kin Vb              &  108      \\ 
\hline
           &   Kin I               &  591      \\
           &   Kin IIa             & 349       \\
 GVI       &   Kin IIb             &  33       \\
           &   Kin IIIa            &  527      \\
           &   Kin IIIb            &  49       \\
\hline
\end{tabular}
\caption{\label{tab:singles} The combined proton and pion singles
rates for the HMS spectrometer. The parameters of the kinematic settings are
given in Table~\ref{tab:kinema}.}
\end{table}
\end{center}

The trigger will be a coincidence between the electrons in the SHMS
and the protons in the HMS. The HMS will detect protons using the
standard detector package. The protons can be identified by
coincidence time-of-flight (TOF). The coincidence  time difference between the two spectrometers will vary from 45~ns to 105~ns as the proton momentum varies from 993 to 494 MeV/c. We plan to run with the SHMS trigger window with width of 60ns and the HMS trigger window width of 20ns. An aerogel detector will not be necessary
in the HMS detector stack.

The combined proton and pion singles
rates have been kept at the level of  $\sim 500~kHz$ or lower to allow a reliable
tracking efficiency calculation. The total singles rates are
presented in Table~\ref{tab:singles}.

\begin{figure}[h]
\begin{center}
\includegraphics[width = 14cm]{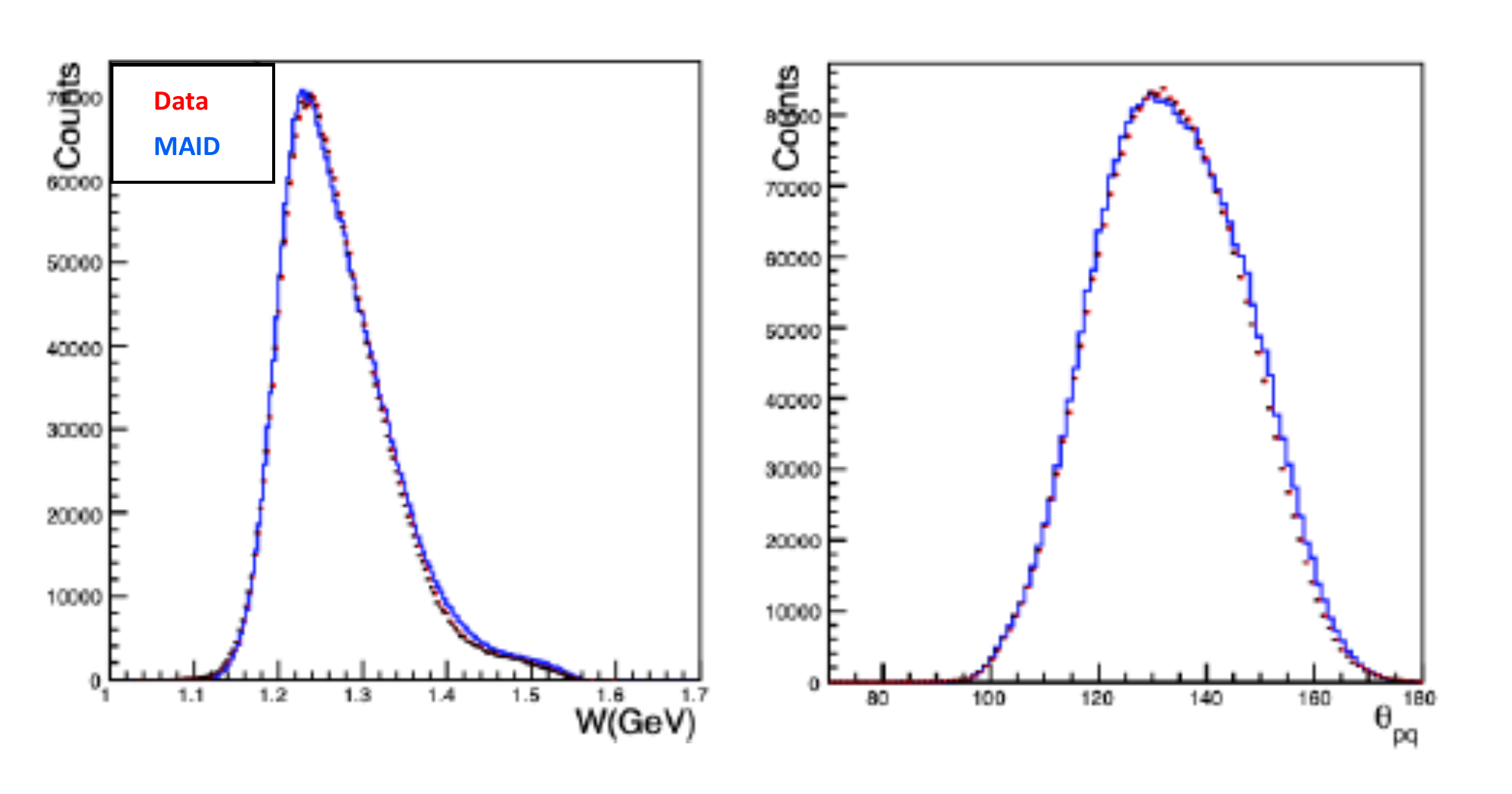}
\end{center}
\caption{\label{fig:pion} Measurement of the $p(e,e'p)\pi^{\circ}$ during the VCS-I experiment. The data (red) are compared to the simulation (blue) weighted with the MAID cross section.}
\end{figure}

Time will be needed for calibrations that will involve luminosity scans, target boiling,
tracking efficiency, electronic deadtime and spectrometer optics. The targets for these studies will be a 10-cm long liquid hydrogen, 10-cm aluminum dummy and optics foil targets. For the elasic $ep$ coincidence measurements, the HMS will be at angles of 50$^{\circ}$ to 63$^{\circ}$ and at momentum between 495 to 994~MeV/c while the SHMS will be at angles of  19$^{\circ}$ to 33$^{\circ}$ and at momentum between 924 to 1797~MeV/c. Conservatively, one day has been set aside for this in the run plan, but if other experiments are running the time would be shared with these experiments. Two additional days of dummy-runs will also be required throughout the experiment.

The spectrometer acceptance will allow to measure the pion electroproduction reaction simultaneously with the VCS. The two reaction channels will be cleanly separated through the reconstructed missing mass spectrum. The cross section for $p(e,e'p)\pi^{\circ}$ is very well known in this region and these measurements will serve as a real-time normalization cross-check during the measurement of the VCS cross section. Fig.~\ref{fig:pion} shows the measurement of the $p(e,e'p)\pi^{\circ}$ during the VCS-I experiment that was conducted using the same experimental setup as the one proposed here. These measurements illustrate the excellent understanding of the spectrometer acceptance in the experiment simulation.

\subsection{Kinematic settings and beam time request}

\begin{figure}[h]
\begin{center}
\includegraphics[width=17cm]{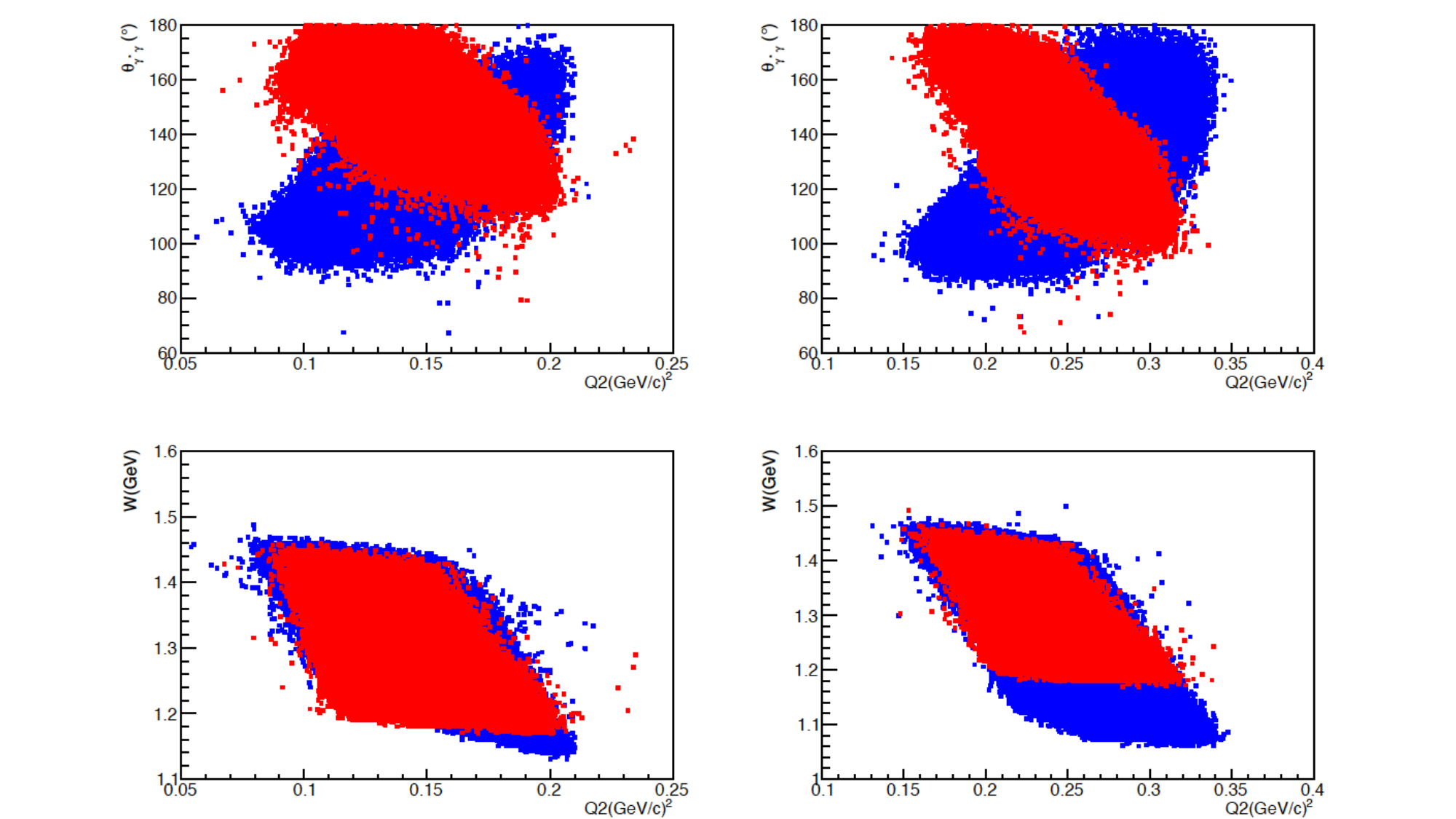}
\end{center}
\caption{\label{fig:phasespace} (Left $Q^{2} =0.15~GeV^{2}$ and Right $Q^{2}=0.25~GeV^{2}$) Correlation of the phase space
variables for a pair of $\phi_{\gamma^*\gamma}=0^\circ$,
$180^\circ$ measurements, with the two different colors
corresponding to $\phi_{\gamma^* \gamma}=0^\circ$ and $180^\circ$
respectively. Left (top and bottom) and right (top and bottom)
correspond to a pair of settings from kinematic groups GII and GIII, respectively.}
\end{figure}

The kinematic settings are summarized in Table~\ref{tab:kinema}.
Monte-Carlo studies have been performed (see
Fig.~\ref{fig:phasespace}) for all of the proposed kinematics using
SIMC~\cite{simc}. The code includes the effects of offsets and
finite resolutions while a physics model is averaged over the finite
acceptances of the experimental apparatus. The reconstructed missing
mass spectrum is presented in Fig.~\ref{fig:q43}. For the
calculation of the count rates and the beam time request the
Dispersion Relations calculation~\cite{gp12} developed by
B.~Pasquini was used and folded over the experimental acceptance.
The model has been proven very successful both in the calculation of
the VCS cross sections above the pion threshold and in the
extraction of the GPs \cite{gp3,gp4,gp5,jlabgp,spaprc}. The beam
time request per setting is summarized at the last column of
Table~\ref{tab:kinema}. The accidental rates have been studied
for all the kinematical settings. The beam current will vary per setting, to a maximum of I=75~$\mu$A, in order to keep the singles rates within the operational range for both spectrometers. The HMS singles rates will range
between approximately 20~kHz and 500~kHz, depending on the setting. The SHMS singles rates will range from 200~kHz to 500~kHz. The coincidence signal to
noise (S/N) ratio will range from $\approx$~7 to 0.3 and
for a wide missing mass cut of $\sim 100~MeV$; a more tight missing
mass applied during the analysis will be able to further improve the
S/N ratio. Calibration data will be taken for normalization studies, while these studies will be further complemented by the simultaneous measurement of the
$p(e,e'p)\pi^\circ$ reaction during the VCS measurements, as commented in the previous section.

\begin{center}
\begin{table} \large
\begin{tabular}{|c|c|c|cc|cc|c|c|}
\hline
Kinematic      & Kinematic          &  $\theta_{\gamma^*\gamma}^{\circ}$    & $\theta_e^{\circ}$ & $P^{'}_{e} (MeV/c)$ &  $\theta_p^{\circ}$  & $P^{'}_{p} (MeV/c)$ & $I~(\mu A)$ & beam time \\
  Group   & Setting              &                                           &                      &                     &                        &                     &     &  (days)   \\
\hline
           &   Kin I               &  $ 110 $           &  $14.3 $             & $ 736.3 $           & $ 54.45 $              & $ 493.93 $     & 15 &  1.00   \\
           &   Kin II              &  $ 133 $           &  $14.3 $             & $ 736.3 $           & $ 44.93 $              & $ 556.10 $     & 15 &  1.00   \\
GI       &   Kin IIIa            &  $ 147 $           &  $14.3 $             & $ 736.3 $           & $ 11.26 $              & $ 583.05 $     & 15 &  1.00   \\
           &   Kin IIIb            &  $ 147 $           &  $14.3 $             & $ 736.3 $           & $ 39.06 $              & $ 583.05 $     & 15 &  1.00   \\
           &   Kin IVa             &  $ 160 $           &  $14.3 $             & $ 736.3 $           & $ 16.73 $              & $ 599.95 $     & 15 &  1.00   \\
           &   Kin IVb             &  $ 160 $           &  $14.3 $             & $ 736.3 $           & $ 33.59 $              & $ 599.95 $     & 15 &  1.00   \\

\hline
           &   Kin I               &  $ 115 $           &  $11.22 $            & $ 1783.0 $          & $ 15.33 $              & $ 615.69 $     & 10 &  1.50   \\
 GII      &   Kin IIa             &  $ 125 $           &  $11.22 $            & $ 1783.0 $          & $ 56.54 $              & $ 647.85 $     & 10 &  2.50   \\
           &   Kin IIb             &  $ 125 $           &  $11.22 $            & $ 1783.0 $          & $ 18.60 $              & $ 647.85 $     & 10 &  1.50   \\
           &   Kin IIIa            &  $ 145 $           &  $11.22 $            & $ 1783.0 $          & $ 49.77 $              & $ 697.99 $     & 10 &  1.50   \\
           &   Kin IIIb            &  $ 145 $           &  $11.22 $            & $ 1783.0 $          & $ 25.37 $              & $ 697.99 $     & 10 &  1.00   \\
           &   Kin IVa             &  $ 165 $           &  $11.22 $            & $ 1783.0 $          & $ 42.82 $              & $ 726.87 $     & 10 &  1.00   \\
           &   Kin IVb             &  $ 165 $           &  $11.22 $            & $ 1783.0 $          & $ 32.32 $              & $ 726.87 $     & 10 &  1.00   \\
\hline
           &   Kin I               &  $ 115 $           &  $14.73 $            & $ 1729.7 $          & $ 20.58 $              & $ 706.89  $     & 30 &  1.75   \\
 GIII      &   Kin IIa             &  $ 130 $           &  $14.73 $            & $ 1729.7 $          & $ 54.89 $              & $ 758.24  $     & 30 &  2.00   \\
           &   Kin IIb             &  $ 130 $           &  $14.73 $            & $ 1729.7 $          & $ 24.78 $              & $ 758.24  $     & 30 &  1.75   \\
           &   Kin IIIa            &  $ 150 $           &  $14.73 $            & $ 1729.7 $          & $ 48.99 $              & $ 808.24  $     & 30 &  1.75   \\
           &   Kin IIIb            &  $ 150 $           &  $14.73 $            & $ 1729.7 $          & $ 30.68 $              & $ 808.24  $     & 30 &  1.75   \\
           &   Kin IVa             &  $ 170 $           &  $14.73 $            & $ 1729.7 $          & $ 42.90 $              & $ 834.12  $     & 30 &  1.00   \\
           &   Kin IVb             &  $ 170 $           &  $14.73 $            & $ 1729.7 $          & $ 36.76 $              & $ 834.12  $     & 30 &  1.00   \\
\hline
           &   Kin I               &  $ 100 $           &  $16.32 $            & $ 1749.3 $          & $ 23.83 $              & $ 664.52 $      & 35 &  1.75    \\
 GIV          &   Kin II              &  $ 120 $           &  $16.32 $            & $ 1749.3 $          & $ 28.01 $              & $ 738.39 $      & 50 &  1.25    \\
           &   Kin IIIa            &  $ 140 $           &  $16.32 $            & $ 1749.3 $          & $ 32.84 $              & $ 795.37 $      & 70 &  1.00   \\
           &   Kin IIIb            &  $ 140 $           &  $16.32 $            & $ 1749.3 $          & $ 53.80 $              & $ 795.37 $      & 70 &  2.00   \\
           &   Kin IVa             &  $ 155 $           &  $16.32 $            & $ 1749.3 $          & $ 36.69 $              & $ 824.46 $      & 70 &  1.50   \\
           &   Kin IVb             &  $ 155 $           &  $16.32 $            & $ 1749.3 $          & $ 49.95 $              & $ 824.46 $      & 70 &  2.50   \\
           &   Kin Va              &  $ 170 $           &  $16.32 $            & $ 1749.3 $          & $ 40.66 $              & $ 840.48 $      & 70 &  1.00    \\
           &   Kin Vb              &  $ 170 $           &  $16.32 $            & $ 1749.3 $          & $ 45.99 $              & $ 840.48 $      & 70 &  1.00    \\
\hline
           &   Kin I               &  $ 100 $           &  $17.72 $            & $ 1676.41 $          & $ 19.75 $              & $ 723.69 $     & 35 &  2.00   \\
           &   Kin II              &  $ 120 $           &  $17.72 $            & $ 1676.41 $          & $ 24.25 $              & $ 808.93 $     & 50 &  1.50   \\
           &   Kin IIIa            &  $ 140 $           &  $17.72 $            & $ 1676.41 $          & $ 29.34 $              & $ 874.74 $     & 70 &  1.50   \\
GV           &   Kin IIIb            &  $ 140 $           &  $17.72 $            & $ 1676.41 $          & $ 51.12 $              & $ 874.74 $     & 70 &  2.00   \\
           &   Kin IVa             &  $ 155 $           &  $17.72 $            & $ 1676.41 $          & $ 33.36 $              & $ 908.37 $     & 70 &  2.00   \\
           &   Kin IVb             &  $ 155 $           &  $17.72 $            & $ 1676.41 $          & $ 47.10 $              & $ 908.37 $     & 70 &  2.00   \\
           &   Kin Va              &  $ 170 $           &  $17.72 $            & $ 1676.41 $          & $ 37.47 $              & $ 926.91 $     & 70 &  1.00   \\
           &   Kin Vb              &  $ 170 $           &  $17.72 $            & $ 1676.41 $          & $ 42.99 $              & $ 926.91 $     & 70 &  1.00   \\
\hline
           &   Kin I               &  $ 120 $           &  $20.45 $            & $ 1623.1 $          & $ 25.31 $              & $ 886.59  $     & 75 &  1.00   \\
GVI           &   Kin IIa             &  $ 140 $           &  $20.45 $            & $ 1623.1 $          & $ 29.91 $              & $ 956.82  $     & 75 &  1.00   \\
           &   Kin IIb             &  $ 140 $           &  $20.45 $            & $ 1623.1 $          & $ 49.81 $              & $ 956.82  $     & 75 &  1.50   \\
           &   Kin IIIa            &  $ 155 $           &  $20.45 $            & $ 1623.1 $          & $ 33.58 $              & $ 992.83  $     & 75 &  1.50   \\
           &   Kin IIIb            &  $ 155 $           &  $20.45 $            & $ 1623.1 $          & $ 46.14 $              & $ 992.83  $     & 75 &  2.00   \\
\hline
\end{tabular}
\caption{\label{tab:kinema} The kinematic settings of the proposed
experiment. A 1100 MeV beam is required for the kinematic group GI (namely for 6 days of beamtime). All other kinematic groups (53 days of beamtime) will require a 2200 MeV beam.}
\end{table}
\end{center}

\begin{figure}[p]
\begin{center}
\includegraphics[width=18.5cm]{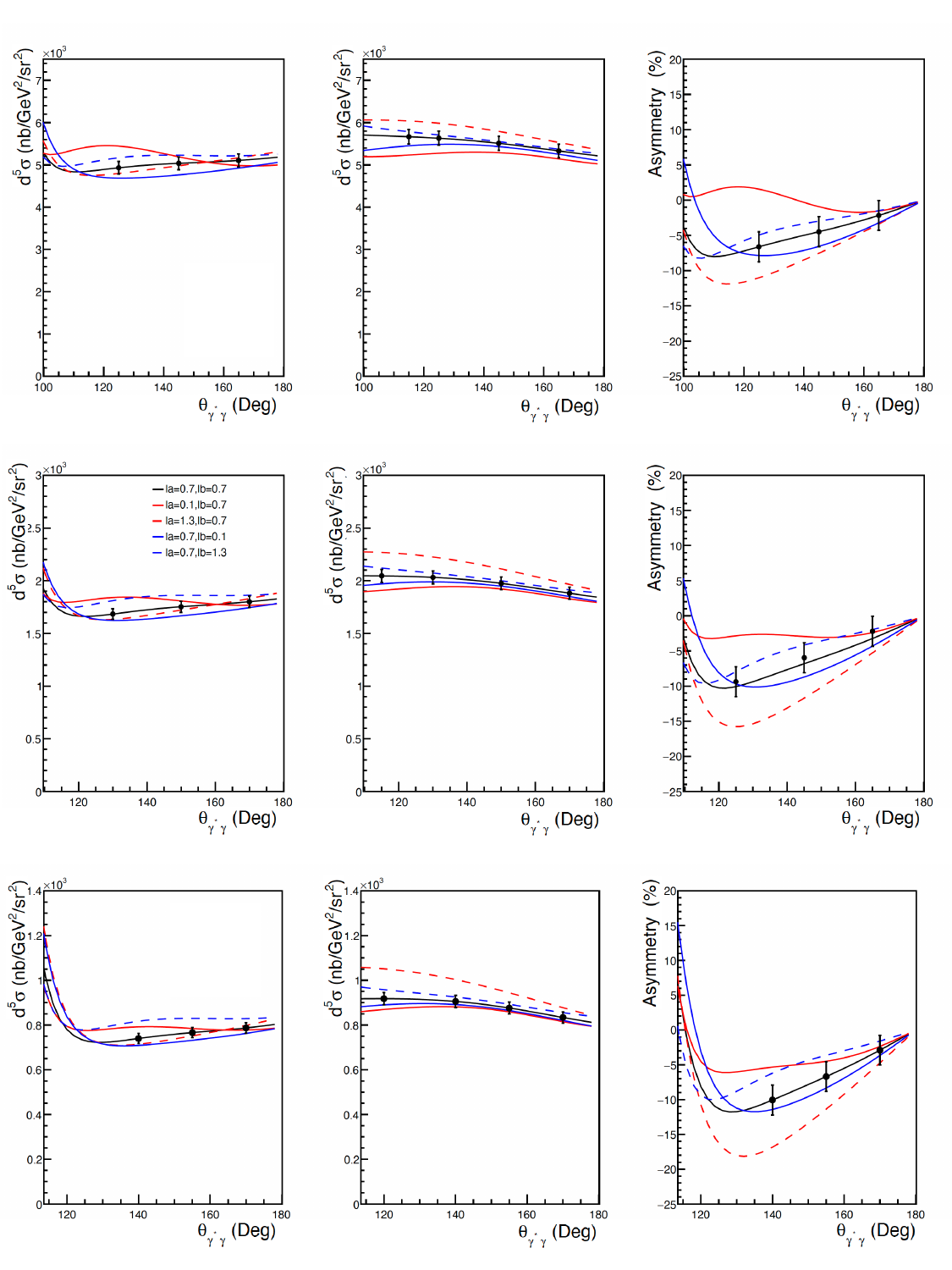}
\end{center}
\vspace{-0.35cm} \caption{From left to right: cross sections at $\phi_{\gamma^* \gamma}=0^\circ$ (left) and $180^\circ$ (center) and asymmetries (right). Top panels correspond to $Q^2=0.15~(GeV/c)^2$, middle panels to $Q^2=0.25~(GeV/c)^2$ and bottom panels to $Q^2=0.35~(GeV/c)^2$.} \label{fig:cross-sec}
\end{figure}

The proposed measurements will allow
the extraction of the GPs at $Q^2=0.05~(GeV/c)^2$, $Q^2=0.15~(GeV/c)^2$, $Q^2=0.25~(GeV/c)^2$,
$Q^2=0.31~(GeV/c)^2$, $Q^2=0.35~(GeV/c)^2$, $0.45~(GeV/c)^2$, and
$0.50~(GeV/c)^2$. All the settings will require a $E_\circ=2.2~GeV$ beam, with the exception of the $Q^2=0.05~(GeV/c)^2$ kinematics that will require a beam energy $E_\circ=1.1~GeV$ The phase space coverage for one pair of asymmetry settings is presented in
Fig.~\ref{fig:phasespace}. 
With the requested beam time, the cross
sections will be measured with a statistical uncertainty ranging
from $\pm1\%$ to $\pm2\%$, depending on the kinematics. 
The systematic uncertainties will be the dominant
factor, ranging at $\approx\pm4\%$. The uncertainty of
the beam energy and of the scattering angle will introduce a
systematic uncertainty to the cross section $\approx\pm2\%$, varying slightly based on the kinematics. 
Other sources of systematic uncertainties involve the target density, target length, beam charge, proton absorption, dead-time, and target cell background; each one of
these contributes $\pm0.5\%$ or less to the uncertainty. The correction for contamination of pions under the photon peak will contribute also at a similar level of $\approx\pm0.5\%$. The uncertainty due to the radiative
corrections will be $\pm1.5\%$. For the tracking efficiencies, we estimate $\pm0.5\%$ (SHMS) and $\pm1\%$ (HMS), based on our recent experience from the analysis of the VCS-I experiment. An uncertainty in the determination of the coincidence acceptance is conservatively accounted for at the $\pm1.5\%$. For the measured asymmetries, the
systematic uncertainties are still larger compared to the
statistical ones, but not as dominant as in the case of the cross
sections. Here, they are expected to be $\approx1\%$ in absolute asymmetry magnitude. 
Other considerations that will contribute towards a better control of calibrations and of systematic uncertainties involve the fact that, with the exception of the first $Q^2$ setting, the beam
energy will remain the same, while the electron spectrometer position and momentum will stay fixed within groups of kinematics.

\begin{figure}[h]
\begin{center}
\includegraphics[width = 16.25cm]{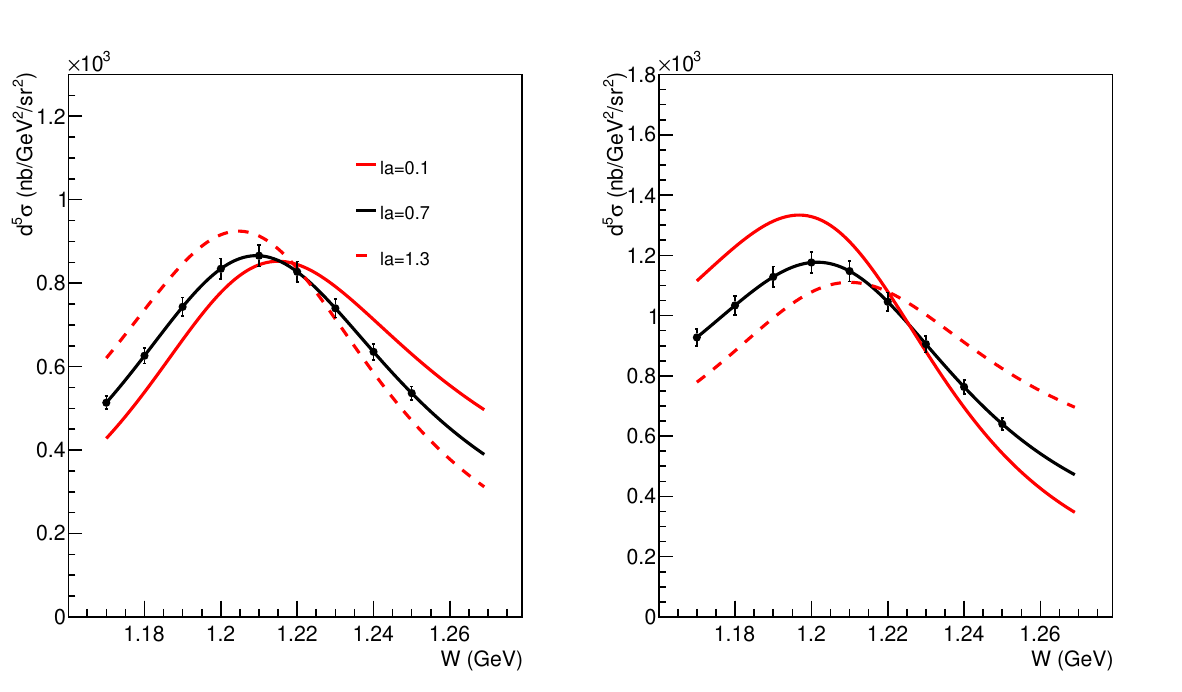}
\end{center}
\vspace{-0.35cm} \vspace{-0.35cm} 
\caption{The W-dependence of the cross section at $Q^2=0.35~(GeV/c)^2$ and $\theta_{\gamma^* \gamma}=140^\circ$, for $\phi_{\gamma^* \gamma}=0^\circ$ (left) and $180^\circ$ (right). The red (solid / dashed) curves illustrate the sensitivity of the VCS cross section to $\alpha_E$.} \label{fig:sensw}
\end{figure}

The extraction of the Generalized
polarizabilities will be performed in a straightforward way through
a fit to the measured cross sections and asymmetries, as was done in
previous measurements~\cite{gp5,jlabgp,gp20,gp21,gp22,gp23}. The mass scale parameters
$\Lambda_\alpha$ and $\Lambda_\beta$ will be fitted by a $\chi^2$
minimization which compares the DR cross sections and asymmetries to
the measured ones, and the two scalar GPs will be determined. The systematic uncertainties will play the leading role in the uncertainties of the extracted electric and the magnetic
GPs. Another source of uncertainty involves the
proton elastic and transition form factors, that enter in the extraction as an input and that are naturally not known with an infinite precision. Here, various parametrizations have been considered and their
impact to the extraction of the scalar GPs has been quantified. This effect is comparable to the level of the statistical uncertainties.




In Fig.~\ref{fig:cross-sec}, projected cross sections and asymmetries
are presented for three different $Q^2$ kinematics and for a fixed $W$ bin. The sensitivity to the electric (and magnetic) generalized polarizability is illustrated through the red (and blue) curves, that correspond to different values for the mass scale parameters $\Lambda_\alpha$ (and
$\Lambda_\beta$, respectively). Improving upon the measurements of the VCS-I experiment, the proposed kinematics extend our physics reach in a targeted way within a region where the sensitivity to the polarizabilities is appreciably changing. In particular for the case of $\alpha_E$, the 
measurements will be conducted within kinematics where the suggested structure (or enhancement) in the polarizability emerges in an anti-diametric way in the VCS cross section. As illustrated in Fig.~\ref{fig:sensw}, the VCS cross section sensitivity to the $\alpha_E$ undergoes a crossing-point and reverses for the two wings of the resonance. Such targeted measurements will allow to largely de-couple the observation of a non-trivial structure in the polarizability from the influence of experimental uncertainties of systematic nature.

\begin{figure}[t]
\begin{center}
\includegraphics[width=13cm]{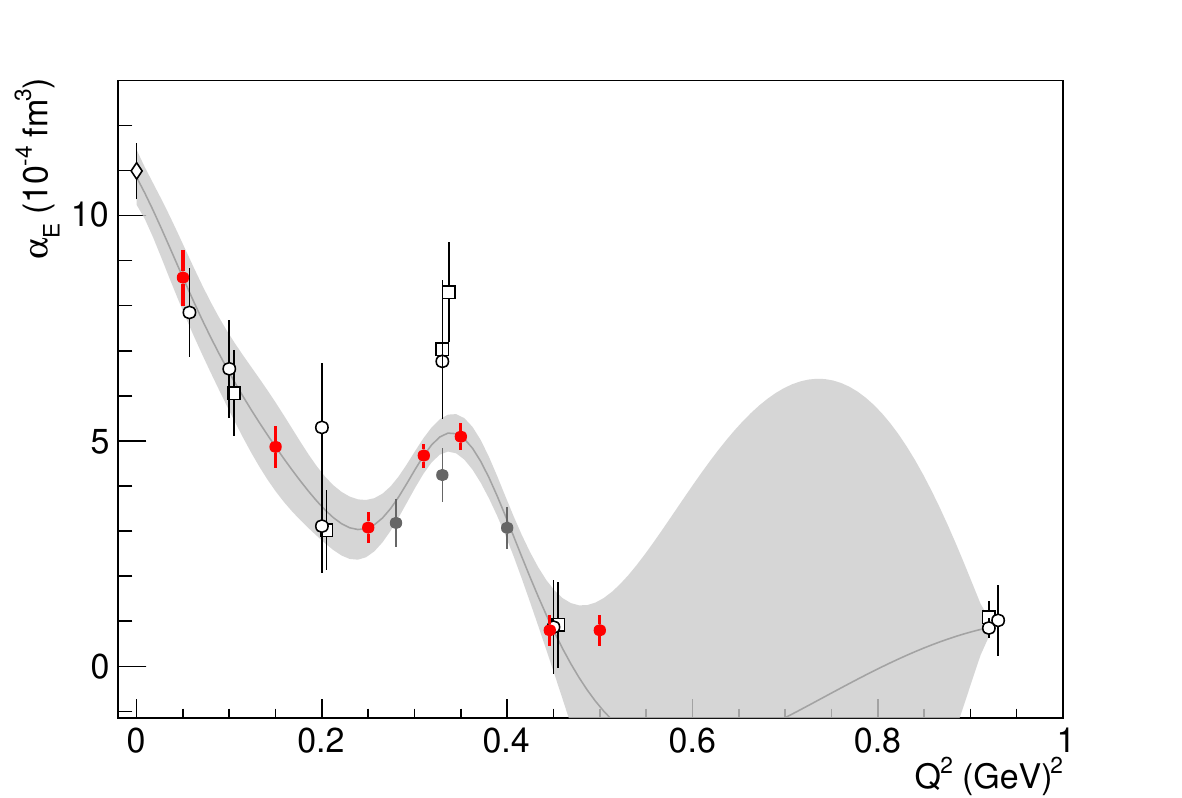}
\end{center}
\vspace{-0.35cm} \vspace{-0.35cm} \caption{The projected
measurements for $\alpha_E$ (red circles). The data points are projected along the extraction of the data-driven GPR method, as shown in Fig.~\ref{fig:alphae}(c). The world data are shown as black points (the VCS-I results are shown with filled gray circles).} \label{fig:aEproj}
\end{figure}

\begin{figure}[h]
\begin{center}
\includegraphics[width=12cm]{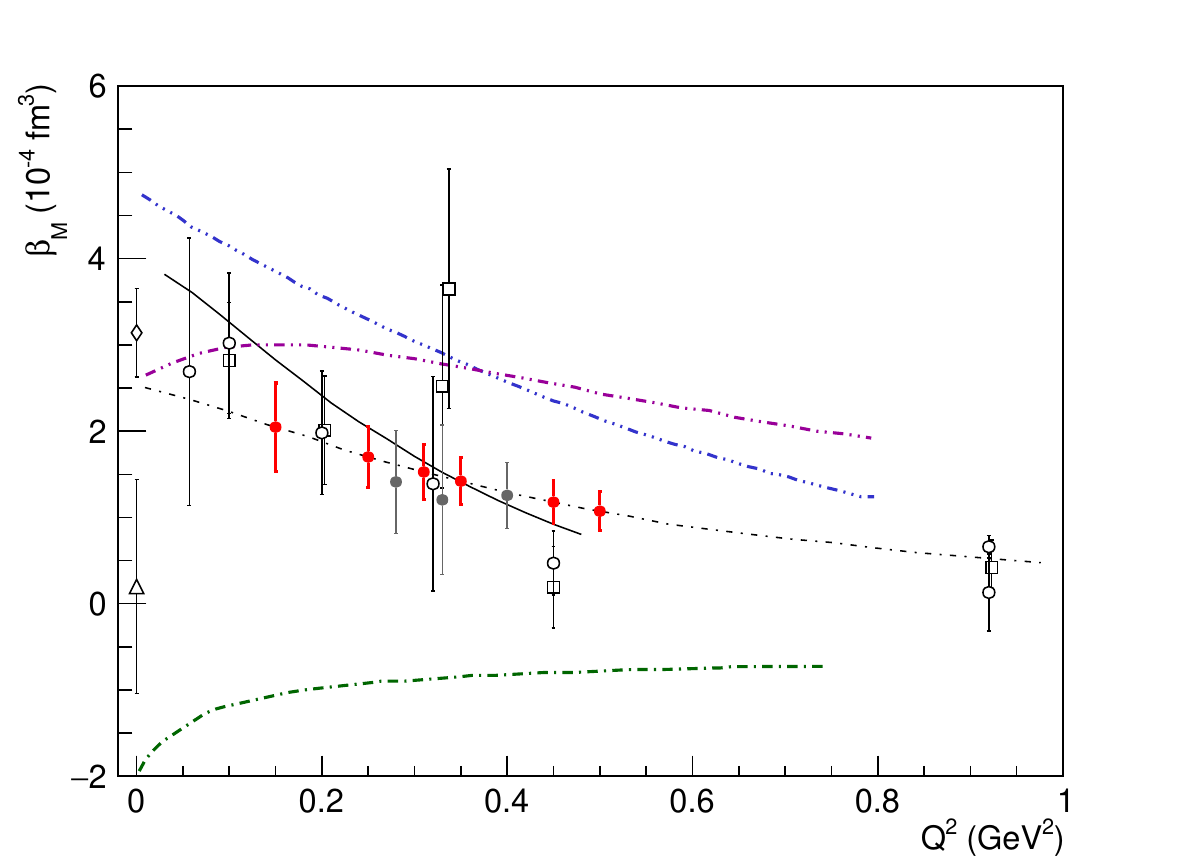}
\end{center}
\vspace{-0.35cm} \vspace{-0.35cm} \caption{The projected
measurements for $\beta_M$ (red circles). Similarly to Fig.~\ref{fig:aEproj}, the world data are shown with black symbols. The theoretical model curves are the same as already referenced in Figs.~\ref{fig:aE-world} and~\ref{fig:betam}.}
\label{fig:bMproj}
\end{figure}

The projected measurements for $\alpha_E$ and
$\beta_M$ are presented in Fig.~\ref{fig:aEproj} and
Fig.~\ref{fig:bMproj}, respectively. The experiment will provide high precision measurements combined with a fine mapping as a function of $Q^2$, lower and higher in $Q^2$ with respect to the kinematics that are of particular interest for $\alpha_E$. They will allow to determine the dynamical signature of $\alpha_E$ through a set of measurements that are of unprecedented precision and that are all consistent in regard to their systematic uncertainties. The measurements will improve significantly the precision in the extraction of the $\beta_M$, providing sufficient information to study the interplay of the paramagnetic and diamagnetic mechanisms in the proton, that is particularly dominant in the low~$Q^2$ region. The measured GPs will allow to extract with high precision the induced polarization and the electric polarizability radius of the proton. The proposed measurements will contribute significantly to our understanding of the nucleon structure by providing stringent tests and guidance to the theory calculations and high-precision benchmark-data for the upcoming Lattice QCD calculations for the proton GPs, that are expected in the near future.

\section{Summary}

We propose to conduct a measurement of the Virtual Compton
Scattering reaction in Hall C, using the HMS and SHMS spectrometers, aiming
to extract the two scalar Generalized
Polarizabilities of the proton in the region of $Q^2=0.05~(GeV/c)^2$
to $Q^2=0.50~(GeV/c)^2$. The experiment will take advantage of the unique capabilities of Hall C, namely
the high resolution of the spectrometers along with the ability
to position them in small angles. In tandem with an extensive measurement of the kinematic phase space, the experiment will pin down the dynamic signature
of $\alpha_E$ and $\beta_M$ with a high precision and a fine mapping as a function of $Q^2$. The proposed measurements will greatly advance our current knowledge of both
$\alpha_E$ and $\beta_M$, which are fundamental quantities of the nucleon that are particularly sensitive to the interplay of the quark and pion degrees of
freedom, and will contribute significantly to our understanding of the nucleon dynamics. {\bf We request a $E_\circ=1.1~GeV$ and $2.2~GeV$ beam at I=75~$\mu$A, a 10~cm liquid hydrogen
target and a total of 59 days of beam on target for the proposed
experiment}: for these 59 days of beamtime, the 53 days will require a beam energy of $E_\circ=2.2~GeV$ and the 6 days will require $E_\circ=1.1~GeV$. Three additional days are requested for optics, dummy, and calibration measurements.



\end{document}